\documentstyle[aps,epsfig]{revtex}
\begin{document}

\draft

\title{A partial wave analysis of the $\pi ^0\pi ^0$ system produced in $\pi ^-p$ 
charge exchange collisions} 

\author{J.~Gunter, A.~R.~Dzierba, R.~Lindenbusch,
K.~A.~Plinske, D.~R.~Rust, E.~Scott,
 M.~R.~Shepherd, P.~T.~Smith, T.~Sulanke, 
 S.~Teige}
\address{\it  Department of Physics, Indiana University,
 Bloomington IN 47405, USA}

\author{A.~P.~Szczepaniak}
\address{\it Nuclear Theory Center, Indiana University,
 Bloomington IN 47405, USA}

\author{S.~U.~Chung, K.~Danyo, R.~W.~Hackenburg, C.~Olchanski,
D. P. Weygand,\footnote[1]{Present address:Thomas Jefferson
National Accelerator Facility, Newport News, VA 23606, USA}
H.~J.~Willutzki}
\address{\it Brookhaven National Laboratory, Upton, NY 11973, USA}

\author{S.~P.~Denisov, V.~A.~Dorofeev, I.~A.~Kachaev, V.~V.~Lipaev, 
A.~V.~Popov, D.~I.~Ryabchikov}
\address{\it Institute for High Energy Physics, Protvino, Russian Federation, 142284} 

\author{Z.~Bar-Yam, J.~P.~Dowd, 
P.~Eugenio,\footnote[2]{Present address:  Department of Physics,
Carnegie Mellon University, Pittsburgh, PA 15213, USA}
M.~Hayek,\footnote[3]{Permanent address: Rafael, Haifa, Israel} 
W.~Kern, E.~King, 
 N.~Shenkav\footnotemark[3]}
\address{\it Department of Physics, University of Massachusetts Dartmouth, 
North Dartmouth, Massachusetts 02747, USA}

\author{ V.~A.~Bodyagin, O.~L.~Kodolova, V.~L.~Korotkikh, 
M.~A.~Kostin, A.~I~Ostrovidov, L.~I.~Sarycheva, 
N.~B.~Sinev, I.~N.~Vardanyan A.~A.~Yershov}
\address{\it Institute for Nuclear Physics, Moscow State University, Moscow, 
Russian Federation 119899} 

\author{D.~S.~Brown,\footnote[4]{Present address: Department of Physics,
University of Maryland, College Park, MD 20742, USA}
 T.~K.~Pedlar, K.~K.~Seth, J.~Wise, D.~Zhao}
\address{\it Department of Physics, Northwestern University, 
Evanston, IL 60208,USA}

\author{ T.~Adams,\footnote[5]{Present address: Department of Physics,
Kansas State University, Manhattan, KS 66506, USA}
 J.~M.~Bishop, N.~M.~Cason, E.~I.~Ivanov, J.~M.~LoSecco,
J.~J.~Manak,
A.~H.~Sanjari ,W.~D.~Shephard, D.~L.~Steinike, 
S.~A.~Taegar,\footnote[6]{Present address: Department of Physics,
University of Arizona, Tucson, AZ 85721, USA}
D.~R.~Thompson}
\address{\it Department of Physics, University of Notre Dame, Notre Dame,
IN 46556} 

\author{G.~S.~Adams, J.~P.~Cummings, J.~Kuhn, J.~Napolitano,
M.~Nozar, J.~A.~Smith, D.~B.~White, M.~Witkowski}
\address{\it Department of Physics, Rensselaer Polytechnic Institute, 
Troy NY 12180,USA}{\small \par}

\author{E852 Collaboration}

\date{\today}
\maketitle
\begin{abstract}
A partial wave analysis  of the $\pi ^0\pi ^0$ system produced in the
 charge exchange reaction: $\pi ^-p\to \pi ^0\pi ^0n$ at an incident momentum
 of $18.3 \ GeV/c$ is presented as a function of ${\pi ^0\pi ^0}$ invariant
 mass, $m_{\pi^0\pi^0}$, and momentum transfer squared, $\left| {t} \right|$,
from the incident $\pi^-$ to the outgoing ${\pi ^0\pi ^0}$ system. For small 
values of $\left| {t} \right|$, the $S$-wave intensity shows a broad 
enhancement at low $m_{\pi ^0\pi ^0}$ with a sharp dip in the vicinity of 
the $f_0\left( {980} \right)$. A dip is also observed in the vicinity of 
the $f_0\left( {1500} \right)$. There is rapid variation of the $S$ to $D_0$ 
relative phase difference in these mass regions. For large values of $\left|
 {t} \right|$, the $f_0\left( {980} \right)$ appears as a bump. 
The $f_2\left( {1270} \right)$, observed in the D-waves, is produced 
dominantly by $\pi$-exchange at low values of $\left| {t} \right|$ and 
$a_2$-exchange at higher values of $\left| {t} \right|$.
\end{abstract}


\section{Introduction}

This paper reports on a high-statistics partial wave analysis (PWA) of the 
$\pi ^{0}\pi ^{0} $ system produced in the charge exchange reaction:
 $ \pi ^{-}p\rightarrow \pi ^{0}\pi ^{0}n $ at an incident momentum of 
$ 18.3\, GeV/c $ using data taken by experiment E852 at Brookhaven National
 Lab (BNL). The PWA was performed over the $ m_{\pi ^{0}\pi ^{0}} $ mass 
range from near threshold ($ 2m_{\pi ^{0}} $) to as high as 
$ 2.2\, GeV/c^{2} $ in $ 0.04\, GeV/c^{2} $ mass bins and in various bins in
 momentum-transfer-squared 
$ t = \left| p_\pi-p_{\pi\pi}\right|^2=\left|p_n-p_p\right|^2$.  

The mass and $\left| t \right|$ dependence of $ \pi \pi $ production in 
$ \pi ^{-} $- induced reactions with one pion exchange (OPE) provides 
information on the process $ \pi \pi \rightarrow \pi \pi $, involving the 
scattering of the lightest hadrons 
\cite{Grayer,Hoogland,Tallahassee,EandM,EandM2,Harada,Kam97}. The extraction of  
$ \pi \pi \rightarrow \pi \pi $ amplitudes is, however,  complicated by the
 presence of production mechanisms other than OPE \cite{Kam97,IM}.  
The $|t|$ and $m_{\pi\pi}$-dependence
 of the partial wave
amplitudes and their relative phases, 
the focus of this paper, provide information on these
mechanisms and the necessary input for
 future $\pi\pi$ scattering studies.

The study of the $\pi\pi$ system also bears on current issues in the 
spectroscopy of conventional $q\bar q$ mesons and non-$q\bar q$ mesons 
such as glueballs or mesonic molecules.  In particular, the isoscalar 
scalar and tensor sectors have more states than can be accommodated within
 the conventional $q\bar q$ model. 
A recent review of light meson spectroscopy \cite{SGJN} includes a
summary of the current experimental situation in these sectors.
Non-$q\bar q$ candidates include the poorly understood 
$f_0(980)$ and the glueball candidates $f_0(1500)$  and $f_J(1710)$,
all of which 
couple to the $\pi\pi$ system \cite{PMWO}. 
Information about the masses,widths, and decay modes of
these states, along with knowledge of their production mechanisms,
as revealed by their $|t|$ dependences, will help in unraveling
their substructure \cite{Ani96,Ani97,Ani97b,Amp87,MRP,EBGR}.   
A complete understanding of these states requires 
corresponding information from $\eta\eta$ and $K\overline{K}$ final states 
as well.  This paper presents information which may be used in such a program.

The $ J^{PC} $ of the $ \pi ^{0}\pi ^{0} $ system
must have $ \emph {J} $ even with both 
$ \emph {P} $ and $ \emph {C} $ positive. The isospin must 
also be even $ (\emph {I = 0} $ or $ \emph {I = 2}) $ for
 $ \pi ^{0}\pi ^{0} $. 
The $ \pi ^{0}\pi ^{0} $ system is thus particularly attractive for investigation 
of scalar and tensor states as
the PWA is simplified without the presence of
 odd angular momenta. 

The $\pi^{-} p \to \pi^{0}  \pi^{0} n$ 
reaction has been studied in experiments with incident $\pi^{-}$ momenta of
$ 9\, GeV/c $ \cite{Fukui},
$ 25\, GeV/c $ \cite{Apel82}, $ 38\, GeV/c $ \cite{Ald95} and $ 100\, GeV/c $
\cite{Ald98}.    The combined information from these
experiments can be used to provide information on how cross
sections of produced states and relative ratios of partial waves
depend on center-of-mass energy.

This paper is organized as follows: The experimental overview is
presented  in Section 2. Event reconstruction and data selection are
described in
 Section 3, where  the general features of the distributions 
in $ m_{\pi ^{0}\pi ^{0}} $ and $|t|$
are also discussed. The details of the PWA formalism and 
results are given in Section 4.  In Section 5 Regge-models are fitted to 
the results from Section 4.  The conclusions are summarized in Section 6.

\section{Experimental Overview}

The E852 apparatus\cite{Te98} was built around and included the Multi-Particle 
Spectrometer (MPS) at BNL. The data used for the analysis reported in this paper were 
collected in 1994
and 1995 using a beam of negatively charged particles of momentum $ 18.3\, GeV/c $. A
30-cm liquid hydrogen target was surrounded by a cylindrical drift chamber
\cite{Baryam97} and
an array of thallium-doped CsI crystals \cite{Adams96}
arranged in a barrel, all located inside the MPS
dipole magnet. Drift chambers were used to track charged particles downstream
of the target. Two proportional wire chambers (PWC's), downstream of the target,
were used in requiring specific charged particle multiplicities in the event
trigger. A 3000-element lead glass detector (LGD) \cite{Cr97} measured the
energies and positions of photons in the forward direction. The dimensions
of the LGD matched the downstream aperture of the MPS magnet. Photons missing
the LGD were detected by the CsI array or by a lead/scintillator sandwich array
(DEA) arranged in a picture frame downstream of the target with an aperture
to allow for the passage of charged particles. 

The first level trigger required that the unscattered or elastically scattered
beam not enter an arrangement of two small beam-veto
 scintillation counters located in front
of the LGD. The next level of trigger required that there be no signal in the
DEA and no charged particles recorded in the cylindrical drift chamber surrounding
the target or in the PWC's (an all-neutral trigger). In the 1994 run, all layers
of the cylindrical drift chamber were used in the trigger requirement, whereas
in the 1995 run, only the outer layer was used.  A common off-line analysis
criterion required no hits in the cylindrical drift chamber. The
final trigger requirement was a minimum deposition of electromagnetic energy
in the LGD.

The LGD is central to this analysis and it is described in detail in reference \cite{Cr97}.
The LGD was initially calibrated by moving each module into a monoenergetic
electron beam. Further calibration was performed by adjusting the calibration
constant for each module until the width of the $ \pi ^{0} $ and $ \eta  $
peaks in the $ \gamma \gamma  $ effective mass distribution was minimized.
The calibration constants were also used for a trigger processor which did
a digital calculation of energy deposited in the LGD and the effective mass
of photons striking the LGD. A laser-based monitoring system allowed for tracking the 
gains of individual modules. 

Studies were made of various algorithms for finding cluster of energies deposited
by photons including issues of photon-to-photon separation and position finding
resolution. These are also described in reference \cite{Cr97}.

\section{Event Reconstruction and Data Selection}
\label{section:reconstruction}
\begin{figure}[htbp]\centering
\begin{tabular}{cc}
\mbox{\epsfig{file=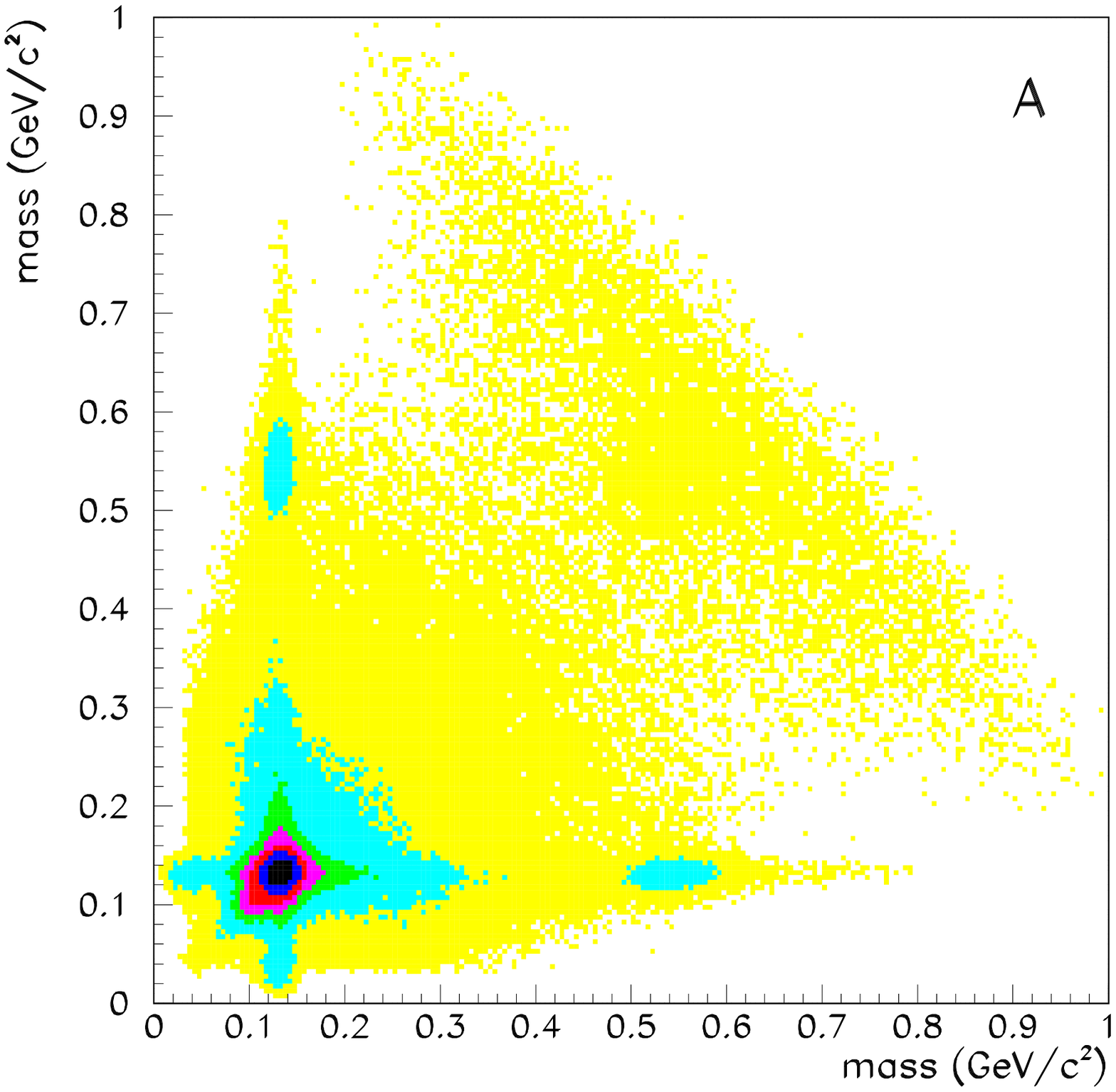,width=0.45\textwidth,height=0.45\textwidth}} &
\mbox{\epsfig{file=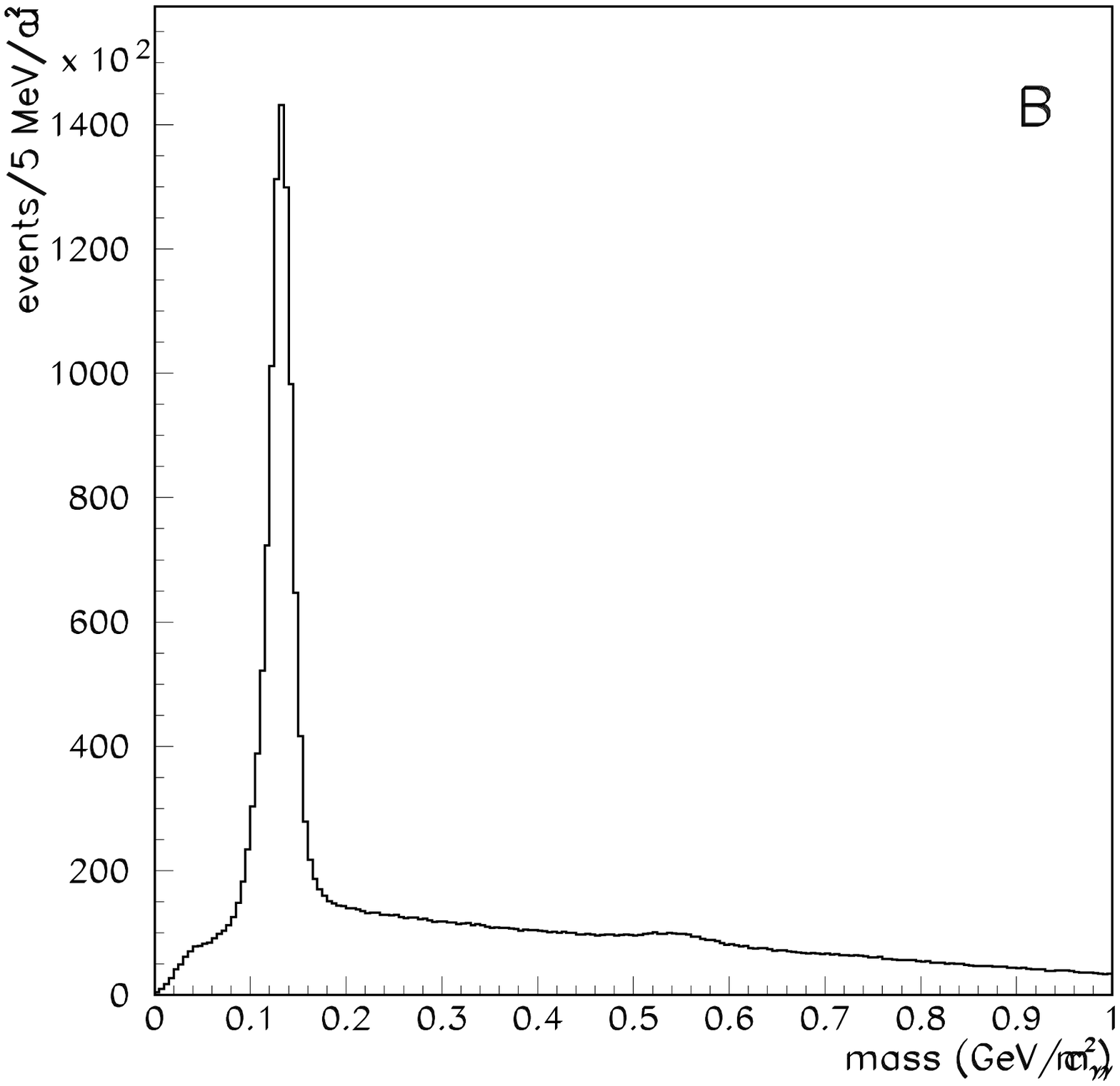,width=0.45\textwidth,height=0.45\textwidth}} \\
\end{tabular}

\caption{\label{m12vsm34}(a) The plot of pairs of di-photon effective masses (\protect$ 
m_{ij}\protect $
vs. \protect$ m_{kl}\protect $) for all pairs of photons (\protect$ i,j,k,l\protect $) is 
dominated by the \protect$ \pi ^{0}\pi ^{0}\protect $ signal. Clear evidence
is also seen for the production of \protect$ \eta \pi ^{0}\protect $. 
(b) The projection of the scatter plot is shown.}

\end{figure}

The combined data sets taken in 1994 and 1995 contain approximately 70 million
all-neutral triggered events. Of these events, approximately 13 million were found
to have four photons in the LGD. The plot of di-photon effective masses for all
possible pairings of photons is shown in the scatterplot of figure \ref{m12vsm34}(a) 
and the projection is shown in  figure \ref{m12vsm34}(b). Events consistent
with the production of two $ \pi ^{0} $'s dominate the scatterplot. 
The $\pi^{0}$ mass resolution is 17 $MeV/c^2$. The sample
of 847,460 $ \pi ^{-}p\rightarrow \pi ^{0}\pi ^{0}n $ events was selected
from the 13 million four photon events by  imposing various analysis criteria.  
It was required that no charged
particles were registered in the MPS drift chambers or the cylindrical drift chamber 
surrounding the
liquid hydrogen target. Any event with a photon within $ 8\, cm $ of the center of the beam 
hole or the
outer edge of the LGD was removed. The $ \chi ^{2} $ returned from kinematic fitting to the 
$\pi p \to \pi ^{0}\pi^{0}n $ 
reaction  hypothesis was required to be less than 9.8 (95\% C.L. for a
three-constraint fit). A further demand was that none   of the
other final state hypotheses considered 
($ \eta \pi ^{0}n $, $ \eta \eta n $) had a better $ \chi ^{2} $.
The final criterion was that the CsI detector registered
 less than 20 MeV, a cut which eliminated events with
low-energy $ \pi ^{0} $'s. The  $\pi^{0} \pi^{0}$ mass resolution
improves from 24 $MeV/c^2$ to 16 $MeV/c^2$ at the mass of the
$K_{s}^{0}$ after kinematic fitting.

Background studies were also  carried out. By selecting
events in a given four photon effective mass region and fitting the associated
scatter plot of di-photon effective mass pairings (similar to figure \ref{m12vsm34}),
 the background of non-$ \pi ^{0}\pi ^{0} 
$events
under the signal was found to be very small. Typical signal to noise ratios determined
by these studies are in the range of 50:1. Monte Carlo studies indicate
that combinatoric background from mis-pairing the reconstructed photons is a
few percent below $ m_{\pi \pi }\sim 0.5\, GeV/c^{2} $ and non-existent at
higher masses. 
These studies are described in more detail in reference \cite{jgth}.

\begin{figure}[htbp]\centering


\epsfig{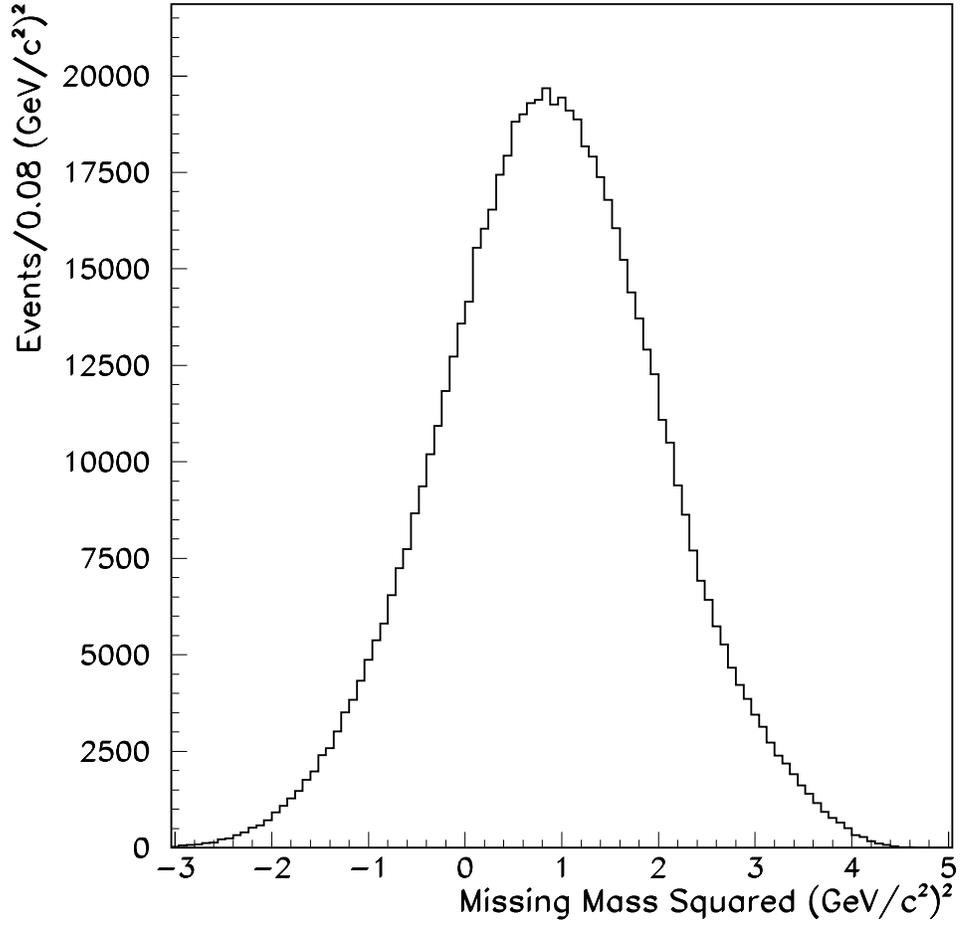}

\caption{\label{fig:mismas} The missing-mass-squared distribution is shown
for four-photon events with a successful kinematic fit to the reaction $\pi^{-} p \to \pi^0 \pi^0 n$}

\end{figure}

\begin{figure}[htbp]\centering
\begin{tabular}{cc}
\mbox{\epsfig{file=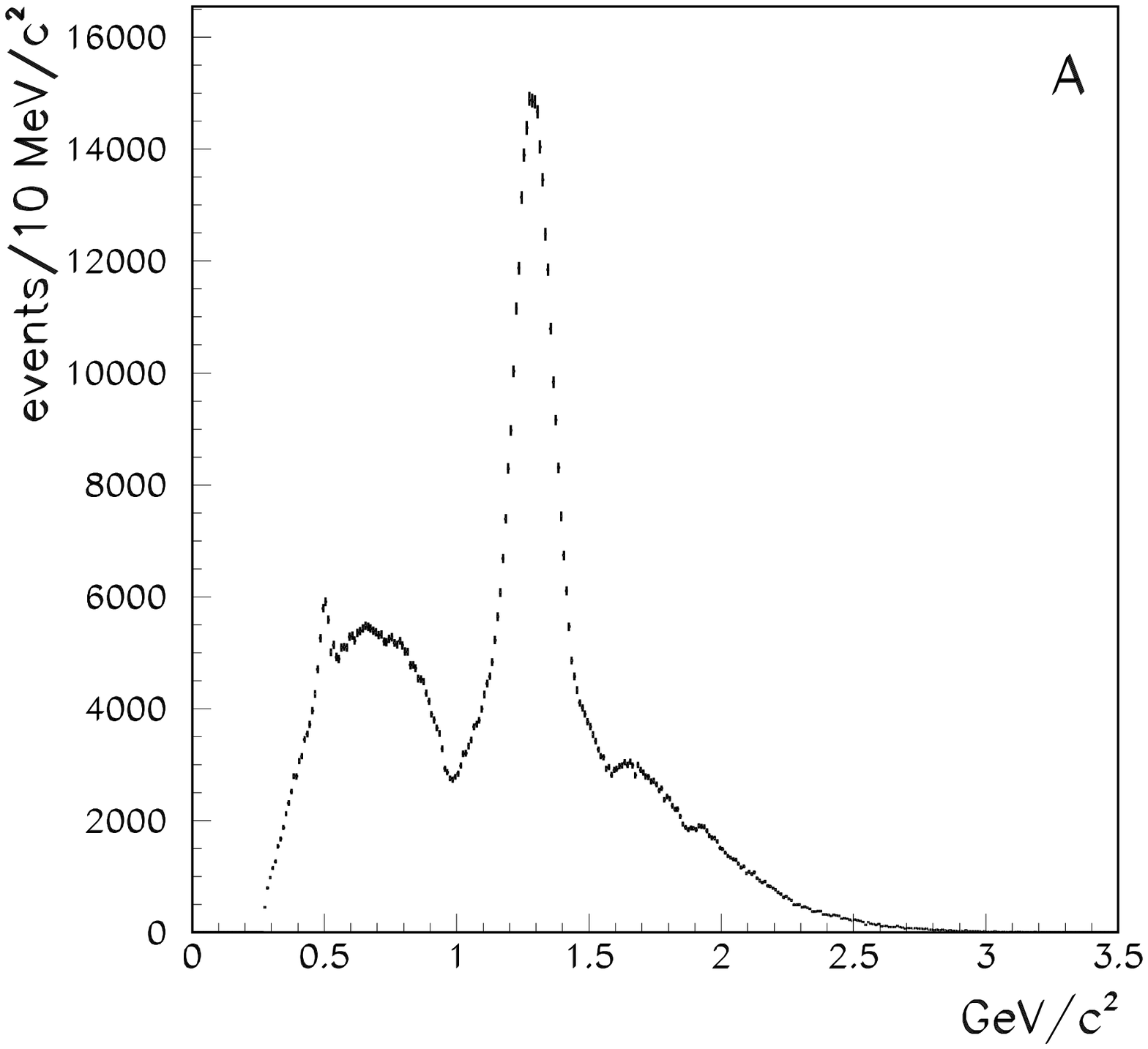,width=0.45\textwidth,height=0.45\textwidth}} &
\mbox{\epsfig{file=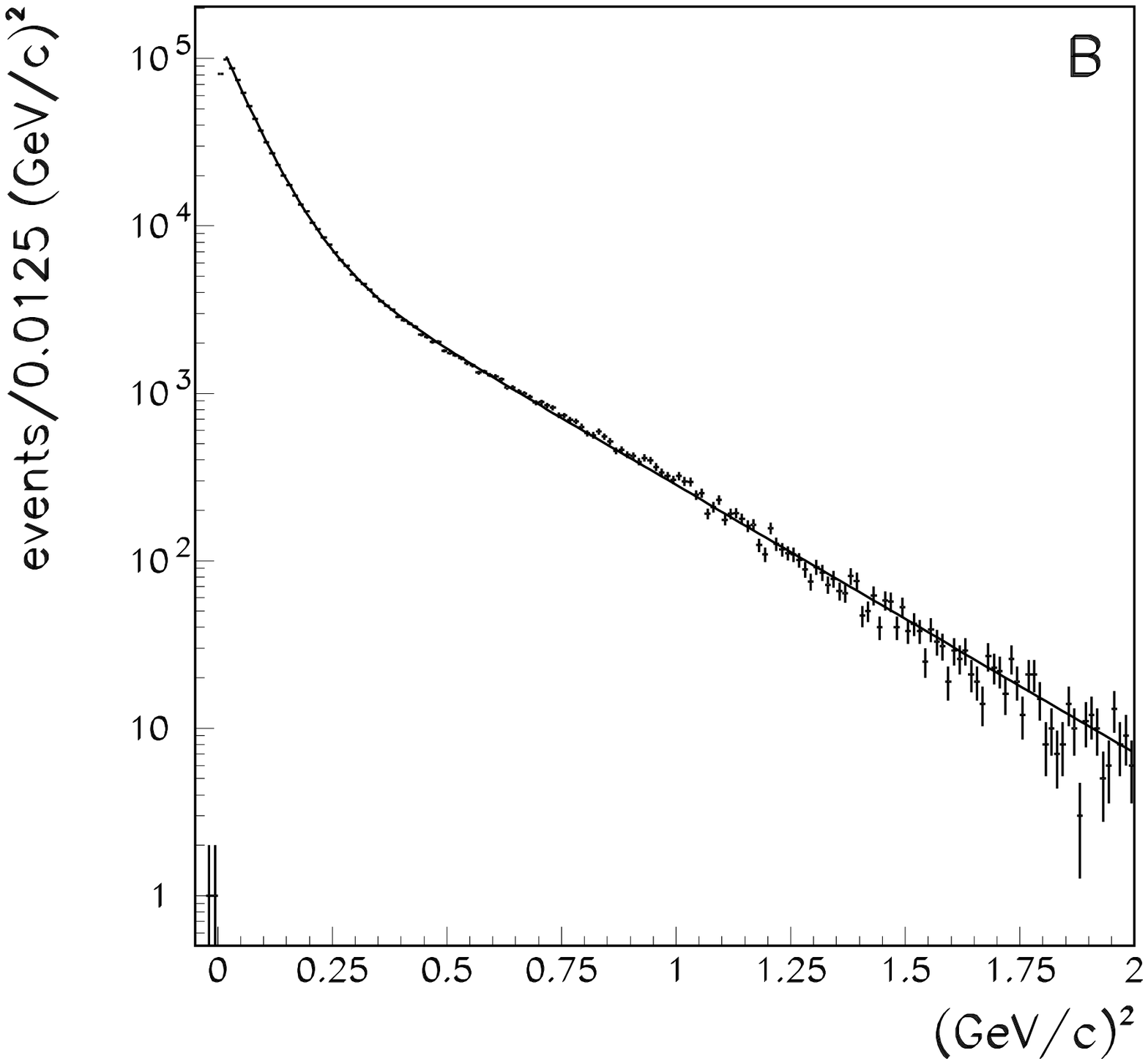,width=0.45\textwidth,height=0.45\textwidth}} \\
\end{tabular}

\caption{\label{fig:m&t}(a) The $\pi^0\pi^0$ effective mass distribution.
This spectrum is dominated by the 
presence of the $f_2(1270)$ resonance.  Additionally, there is a broad enchancement 
peaking near 0.8 $GeV/c^2$ and dips in the spectrum at $1.0$ and $1.5 \ GeV/c^2$.  
(b) The 
momentum-transfer-squared distribution with a fit to the sum of two exponential functions.  
The structure of this distribution is suggestive of changing production mechanisms.}

\end{figure}

The distribution in missing-mass-squared, recoiling against the four photons,
for events with a successful kinematic fit to the reaction
$\pi^{-} p \to \pi^{0} \pi^{0} n$ is shown in figure \ref{fig:mismas}. The 
missing-mass-squared is determined from photon position and energy information
before kinematic fitting and the distribution peaks near the square of
the neutron mass.
The distribution in $ \pi ^{0}\pi ^{0} $ effective mass is shown in figure \ref{fig:m&t}. The 
spectrum is dominated by the $ f_{2}(1270) $ resonance and a broad enhancement at low $ 
\pi ^{0}\pi ^{0} $ mass (from threshold to about $ 1.0\, GeV/c^{2} $). There is also a 
small $K_S^{0}\to \pi ^{0}\pi ^{0}$ signal present, despite the requirement that the 
deposited energy in the CsI detector not exceed $ 20\, MeV $. This CsI energy cut
reduces a substantial fraction of $K_S^{0}$ events but other
reactions producing $K_S^{0}$ can avoid deposition of energy in the CsI detector.
By correlating the observed yields of $K_S^{0}$ and $ f_{2}(1270) $ mesons,
for samples with and without the CsI detector energy cuts, with cross sections for
$ f_{2}(1270) $ production and associated  $K_S^{0}$ production 
($\pi^{-} p \to K_S^{0} \Lambda (\Sigma^{0})$) measured in other experiments, we
estimate an overall CsI detector inefficiency of  $5\%$.  These studies also
indicate that the background level of non-neutron events under the  $ f_{2}(1270) $
is approximately
 $1\%$.
Another feature of the 
spectrum is the dip at $ 1.0\, GeV/c^{2} $, 
which will  be seen to be due to the interference of a narrow
resonance, the $ f_{0}(980) $,  with a broad $ \pi ^{0}\pi ^{0} $ enhancement. 

\begin{figure}[htbp]\centering
\begin{tabular}{cc}
\mbox{\epsfig{file=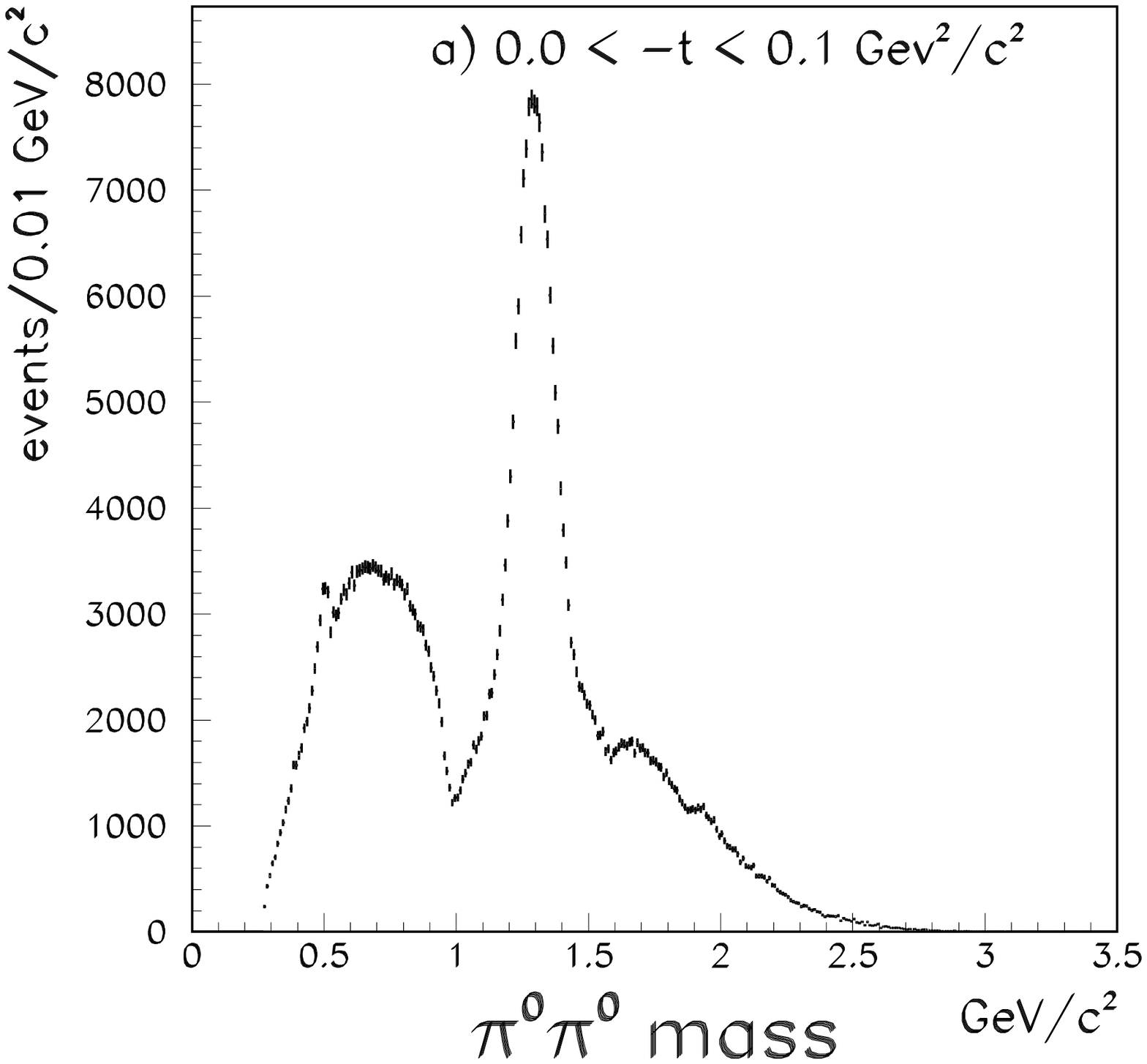,width=0.45\textwidth,height=0.45\textwidth}} &
\mbox{\epsfig{file=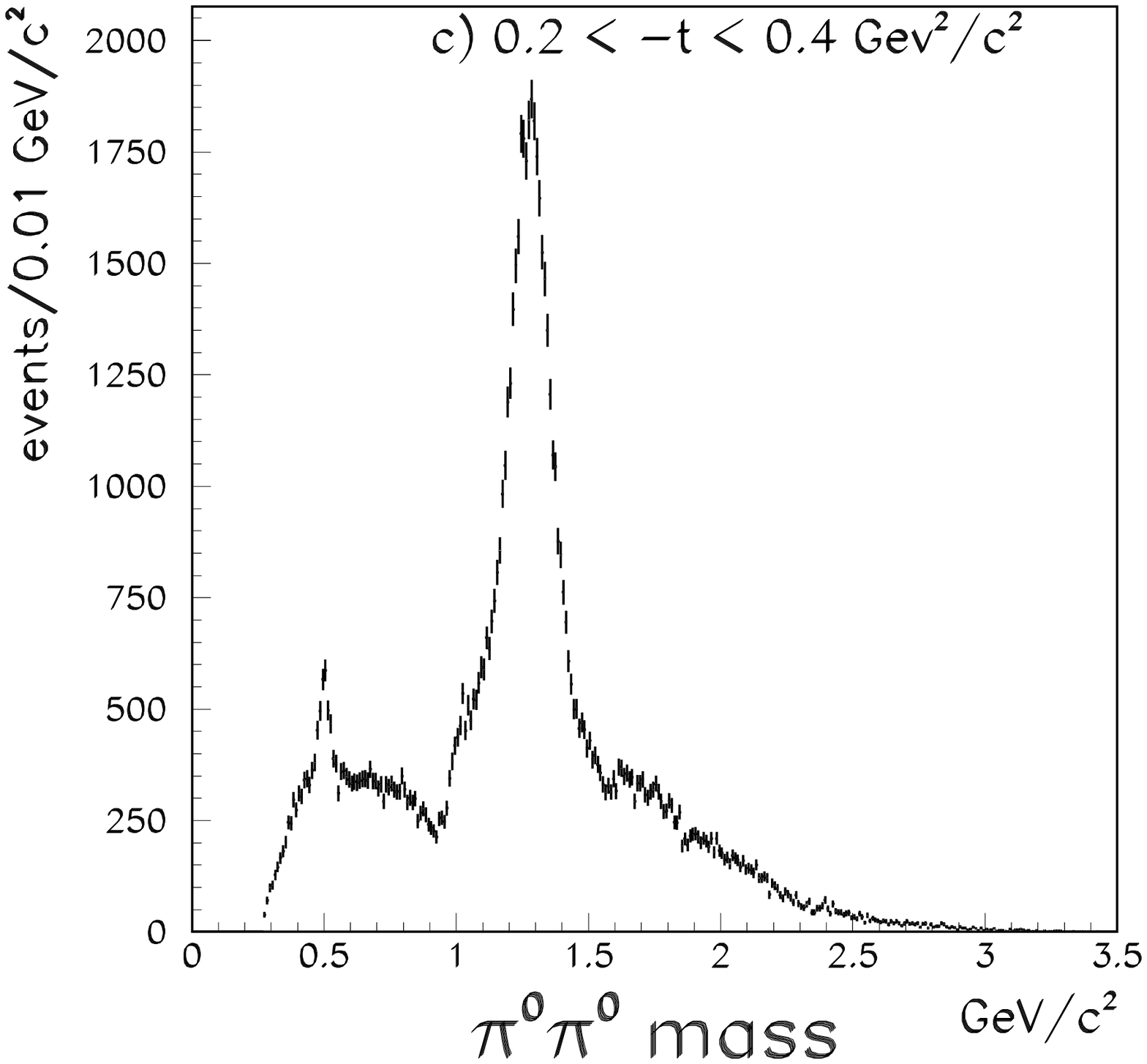,width=0.45\textwidth,height=0.45\textwidth}} \\
\mbox{\epsfig{file=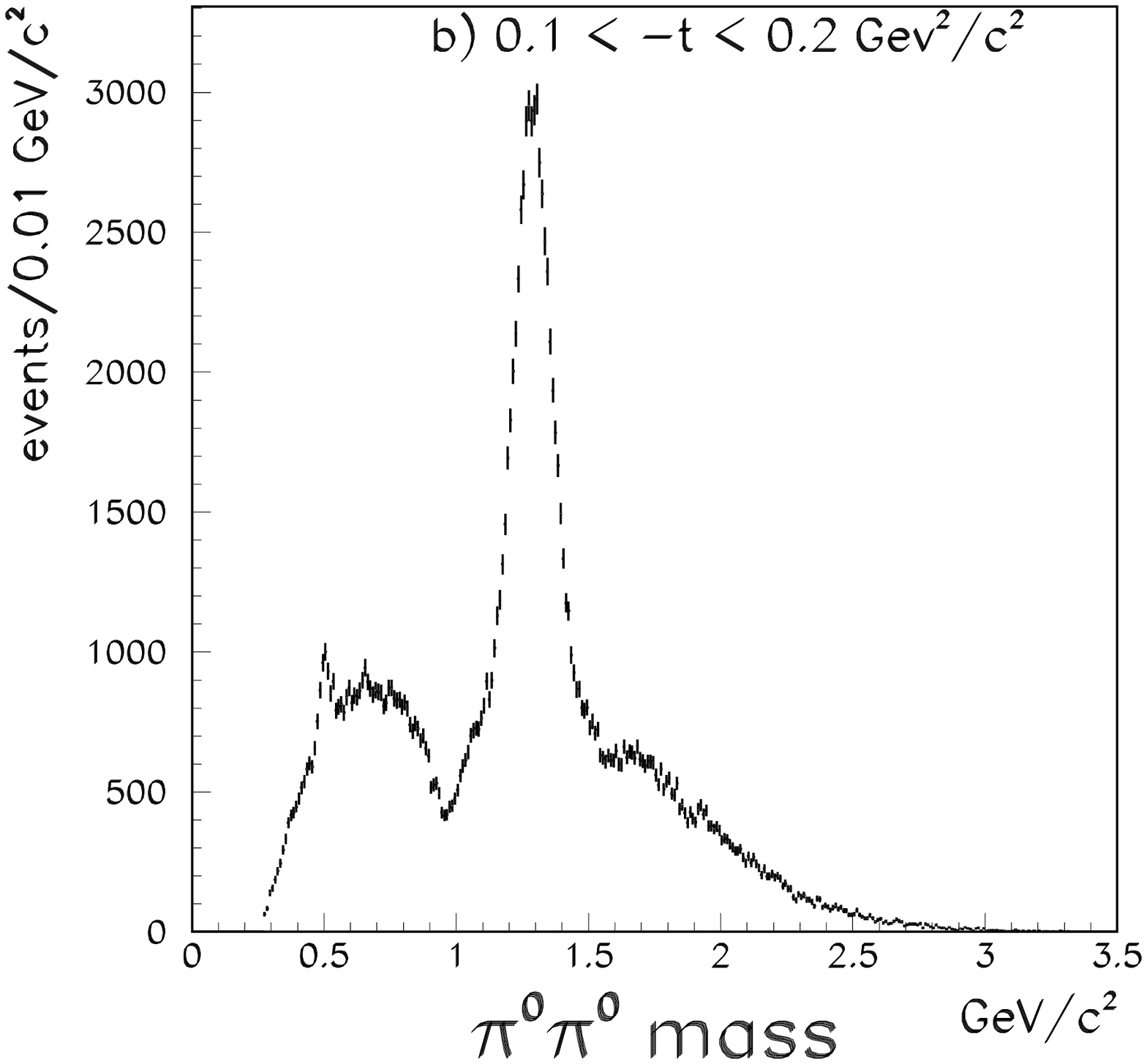,width=0.45\textwidth,height=0.45\textwidth}} &
\mbox{\epsfig{file=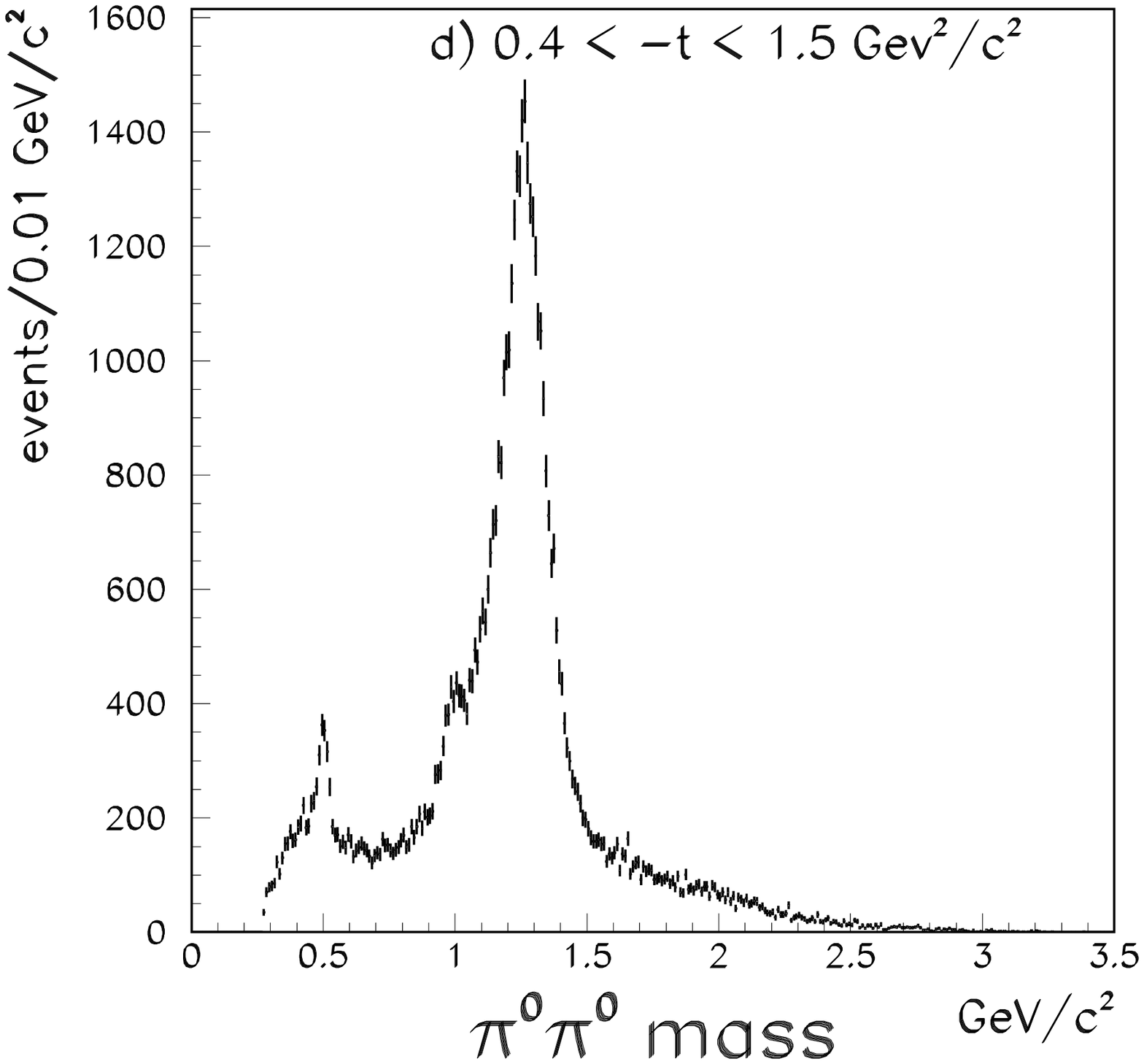,width=0.45\textwidth,height=0.45\textwidth}} 
\end{tabular}

\caption{\label{fig:mAsFofT}The $\pi^0\pi^0$  effective mass distribution for four 
regions of $|t |$. The
shape is seen to be strongly dependent on $| t |$. Of particular interest is the disappearance of the 
broad enhancement near $0.8 \, GeV/c^2$ in (a) and (b) and the emergence of a 
small peak at $0.98 \  GeV/c^2$ in (c) and (d) with increasing values of $|t|$.}

\end{figure}

The distribution in $| t |$, shown in figure \ref{fig:m&t}, is not characterized by a
single exponential, suggesting more than one  
production mechanism. The curve is a fit of this distribution to a sum of two exponentials: 
$dN/dt=a\cdot e^{-b\cdot |t|}+c\cdot e^{-d\cdot |t|}$ where $b=15.5\ \left( {GeV/c} 
\right)^{-2}$ and $d=3.7\ \left( {GeV/c} \right)^{-2}$. 
Based on this structure, we initially 
examine the $ \pi ^{0}\pi ^{0} $ effective mass spectra in four bins in $ |t| $ as shown in 
figure \ref{fig:mAsFofT}.  The $t$-dependence of the $S$, $D_0$, and $D_+$ partial waves 
is later investigated in a set of partial wave fits more finely binned in $|t|$.

An inspection of figure \ref{fig:mAsFofT} reveals striking differences in the $\pi^0\pi^0$ 
mass spectra associated with the four bins in $|t|$. For example, the low-mass structure 
which dominates in figure \ref{fig:mAsFofT}a is much less prominent in 
figure \ref{fig:mAsFofT}d. The dip associated with the $ f_{0}(980) $ resonance in 
figure \ref{fig:mAsFofT}a becomes a bump in figure \ref{fig:mAsFofT}d. These and other 
features are explored in more detail below in the discussion of the PWA results.

\section{Partial Wave Analysis}

Partial wave analysis is used to extract production amplitudes (partial waves)
from the observed decay angular distributions of the di-pion system.  A  
process such as $\pi^-p\rightarrow \pi^0 \pi^0 n $, dominated by $t$-channel meson exchange, 
is  simplest to analyze in the Gottfried-Jackson
reference frame. The Gottfried-Jackson frame is defined as a right-handed coordinate
system in the center of mass of the produced di-pion system with the $ z-axis $
defined by the beam particle momentum and the $ y-axis $ perpendicular
to the plane defined by the beam and recoil neutron momenta. The decay angles
($ \theta ,\phi  $) are determined for one of the produced $ \pi ^{0} $
momenta. At fixed beam momentum, an event is fully specified by $ (m_{\pi \pi },t,\theta 
,\phi ) $.
The data are binned in $ m_{\pi \pi } $ and $ t $ and the
production amplitudes, and their relative phases, are extracted from the
accumulated angular distributions using  an extended maximum likelihood fit to the 
distributions in ($\theta$,$\phi$) \cite{Chu97}.  The naming convention for the partial 
waves is summarized in Table \ref{tab:nomenclature}.

 The explicit form of
the angular distribution $ I(\theta ,\phi ) $ fitted to the data in a given
mass and momentum transfer range in this analysis is given by

\begin{eqnarray}
I(\theta ,\phi )= & \left|S + \sqrt{5}D_{0}P^{0}_{2}(\cos \theta
)-\sqrt{\frac{5}{3}}D_{- }P^{1}_{2}(\cos \theta )\cos \phi
+\sqrt{9}G_{0}P_{4}^{0}(\cos \theta )\right|^{2} 
\nonumber\\
& +\left|\sqrt{\frac{5}{3}}D_{+}P_{2}^{1}(\cos \theta )\sin \phi \right|^{2}  
\end{eqnarray}

where $P^m_l(\theta)$ are the associated Legendre polynomials \cite{Chu97}.

As summarized in Table \ref{tab:nomenclature}, the $D_+$ wave is produced by
the exchange of a particle with natural parity ($P \ = \ (-1)^J$).  For production
of a $\pi \pi$ system, the dominant natural parity
exchange particle is the $ a_{2} $ \cite{IW}.
The $S$, $D_0$, $D_-$ and $G_0$ waves are produced by the exchange of a particle
with unnatural parity ($P \ = \ (-1)^{J+1}$). Again, for $\pi \pi$ production,
the dominant unnatural parity exchange particles are the $ \pi  $ and the
 $ a_{1} $ \cite{IW}.

\subsection{Ambiguities\label{sec:ambigs}}

There are multiple discrete sets of partial wave amplitudes which can give rise
to exactly the same angular distribution \cite{Chu97}.  It can be shown that in a partial 
wave fit with only
$ S,D_{0},D_{-,} $ and $ D_{+} $ partial waves  there are four sets
of ambiguous partial wave amplitudes. The four sets can be divided into two groups with 
different partial wave intensities. Additionally, within each group, there is a sign ambiguity 
in the phases between the amplitudes.   

Normally there are two ambiguities; if one wave in each naturality 
is fixed \emph{a priori}, e.g. set real or to
some complex value for dynamical reasons, there is still an 
overall sign ambiguity. However, even
this sign ambiguity could be fixed by the requirement, for example, 
of a resonant behavior in one of the
waves.

In general, there is an eightfold ambiguity for a $\pi^{0} \pi^{0}$ system 
containing L=0, 2 and 4. However, these
ambiguities necessarily entail nonzero $G_{-}$ and $G_{+}$ waves. 
In this paper we have assumed that these are
negligibly small and searched for ambiguities with nonzero $G_{0}$ wave.
We find no such ambiguities in our data.

In the analysis of the $ \pi ^{0}\pi ^{0} $ system, 
the
physical solution can be selected by a combination of physical arguments
(which will be given below)
and the requirement that solutions be smoothly connected 
as a function of mass. This selection of the physical solution
applies simultaneously to all intensities and phases.
In what follows, 
the physical solution is plotted with solid symbols. The other solutions are plotted with 
open symbols and are presented for completeness.

\subsection{Partial Wave Fits }

\subsubsection{Results for \protect$ 0.01<-t<0.10\, GeV^{2}/c^{2}\protect $ }

\begin{figure}[htbp]\centering
\begin{tabular}{cc}
\mbox{\epsfig{file=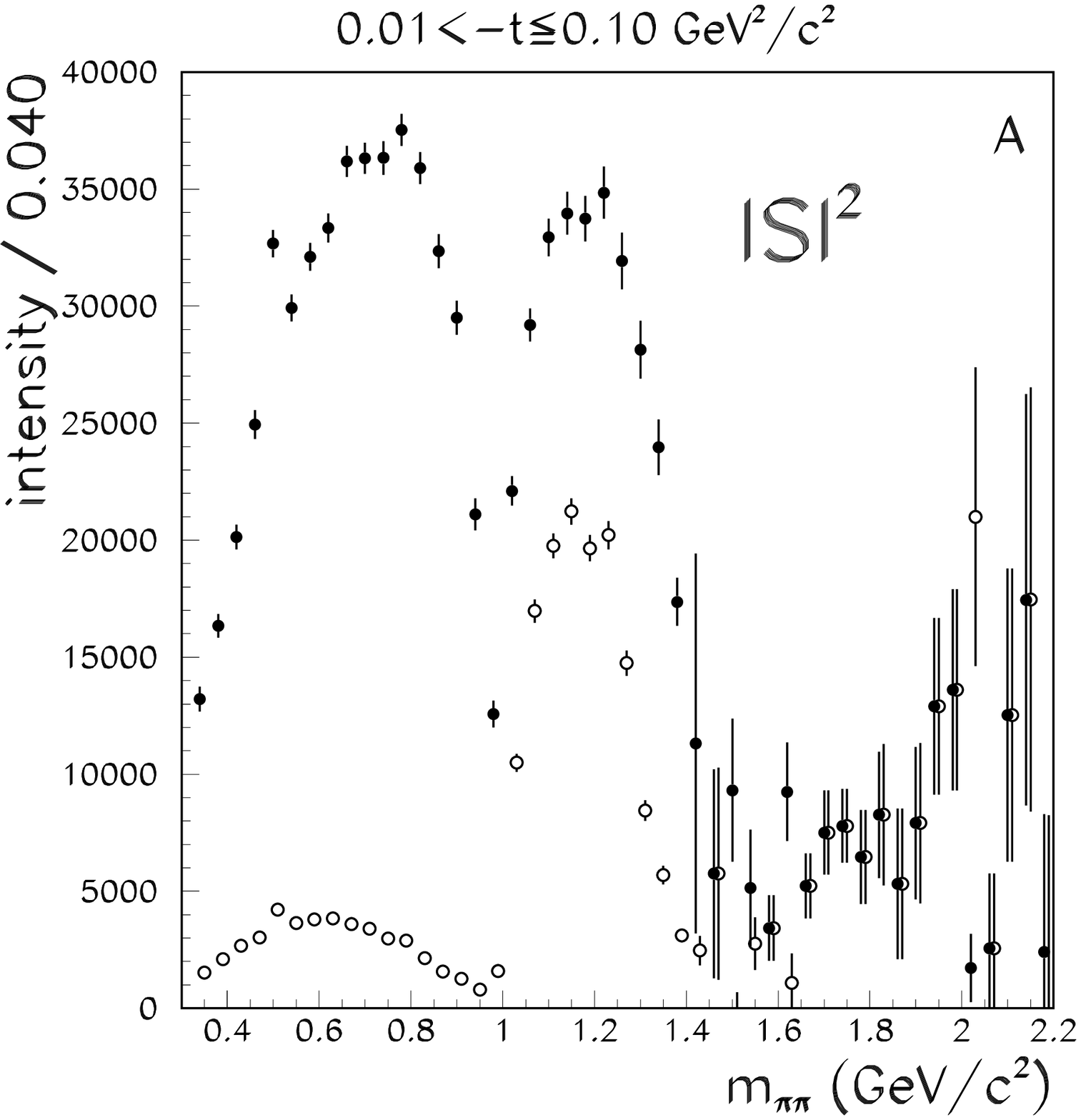,width=0.45\textwidth,height=0.45\textwidth}} &
\mbox{\epsfig{file=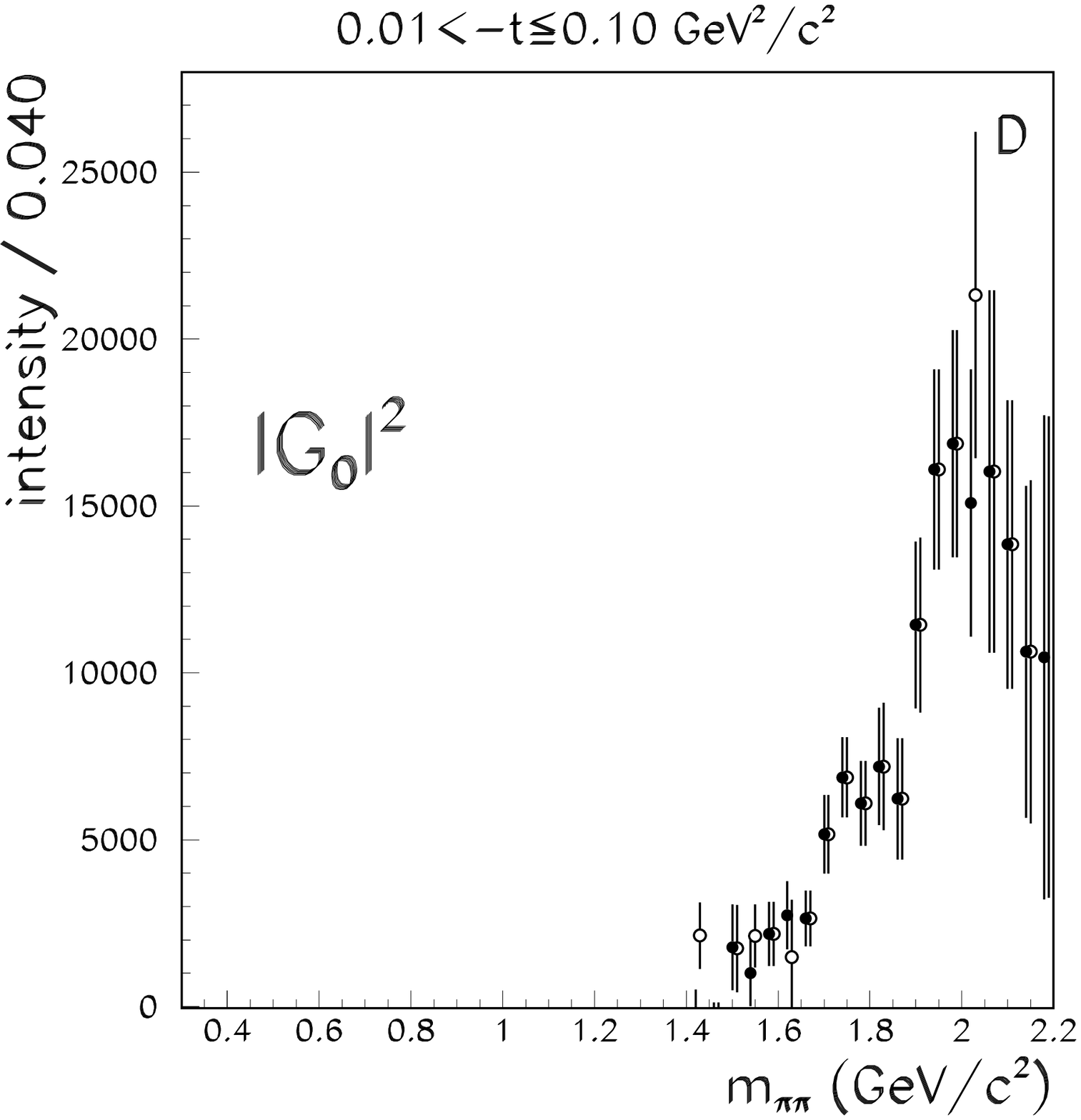,width=0.45\textwidth,height=0.45\textwidth}} \\
\mbox{\epsfig{file=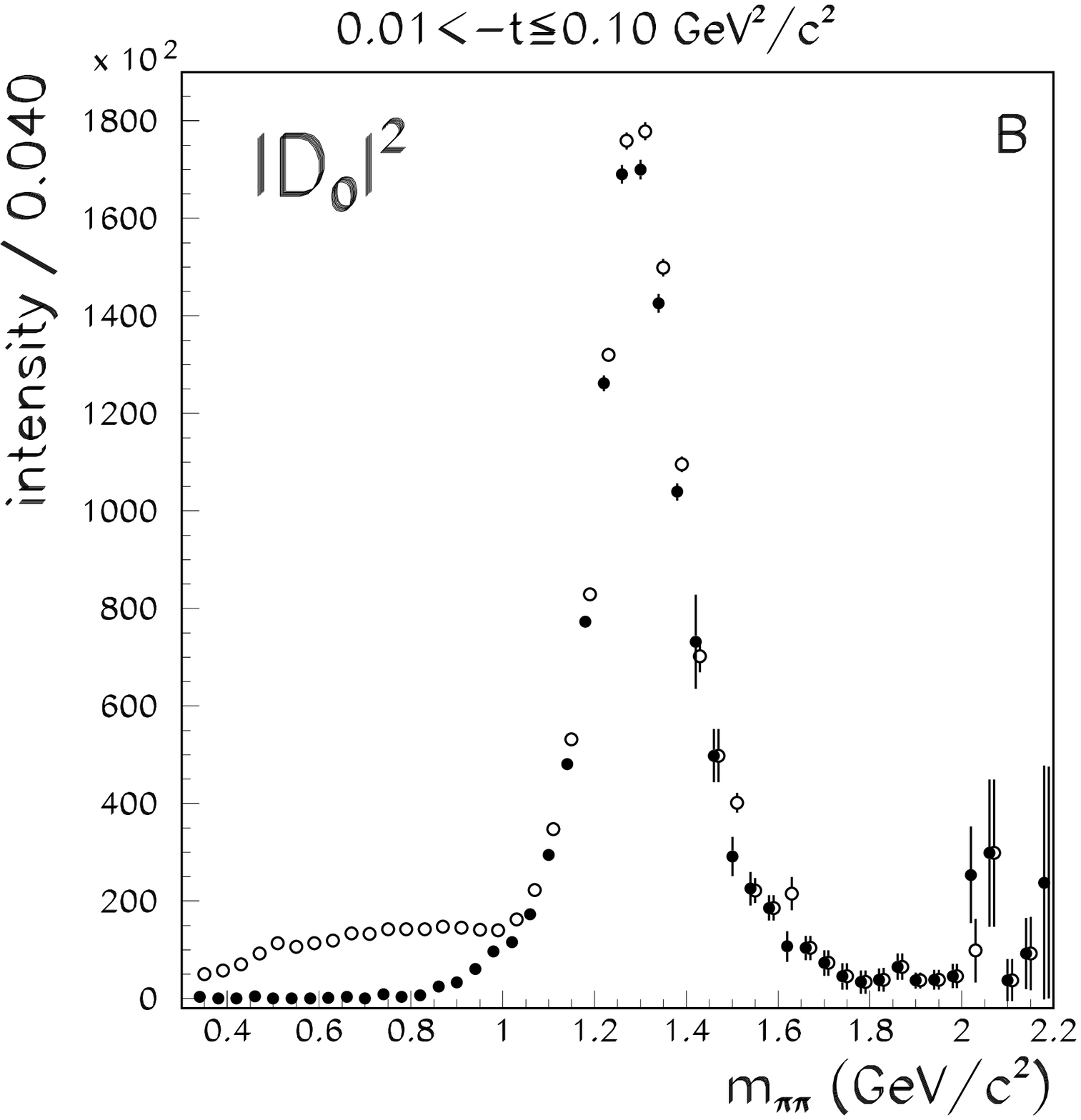,width=0.45\textwidth,height=0.45\textwidth}} &
\mbox{\epsfig{file=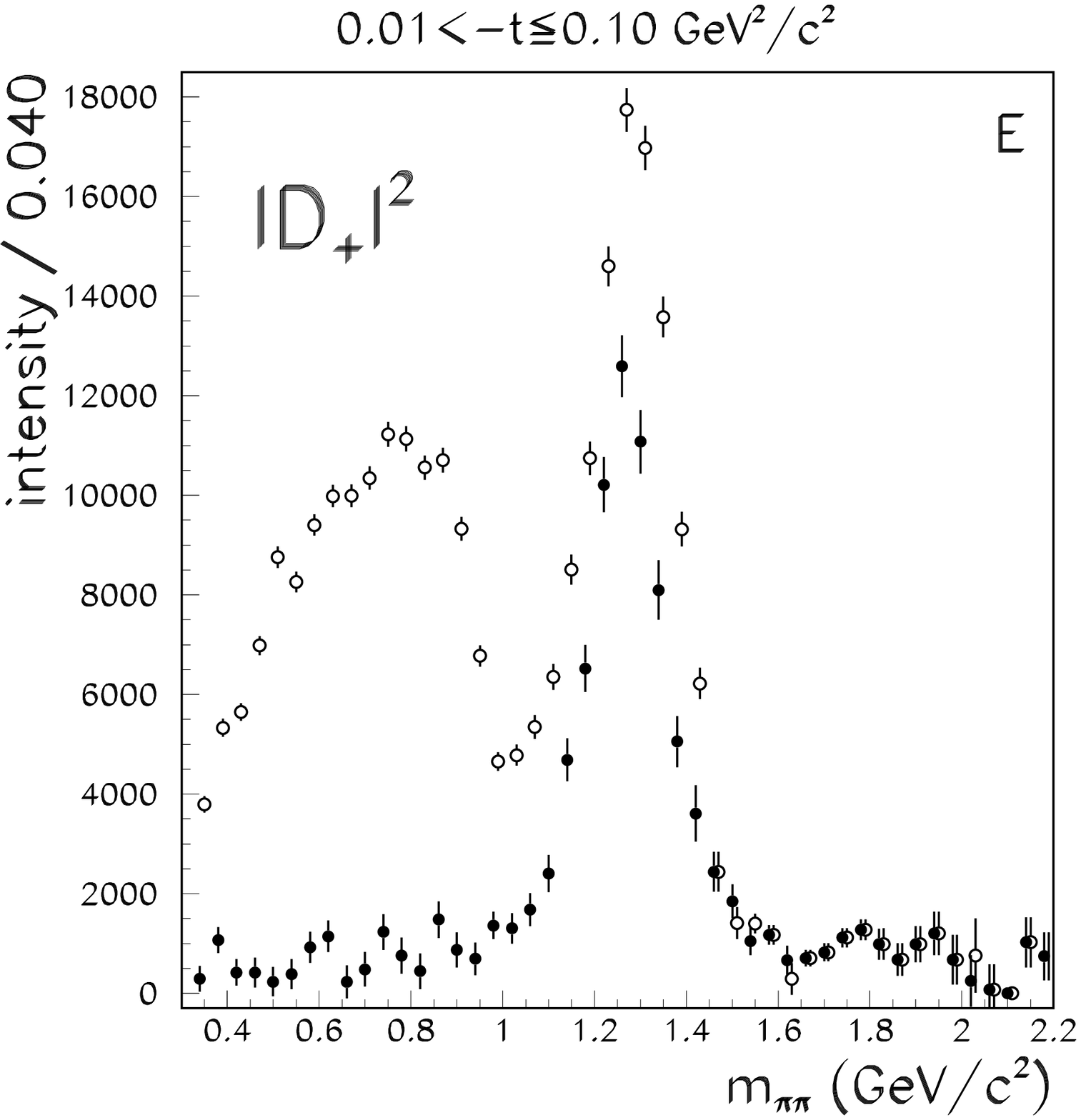,width=0.45\textwidth,height=0.45\textwidth}} \\ 
\mbox{\epsfig{file=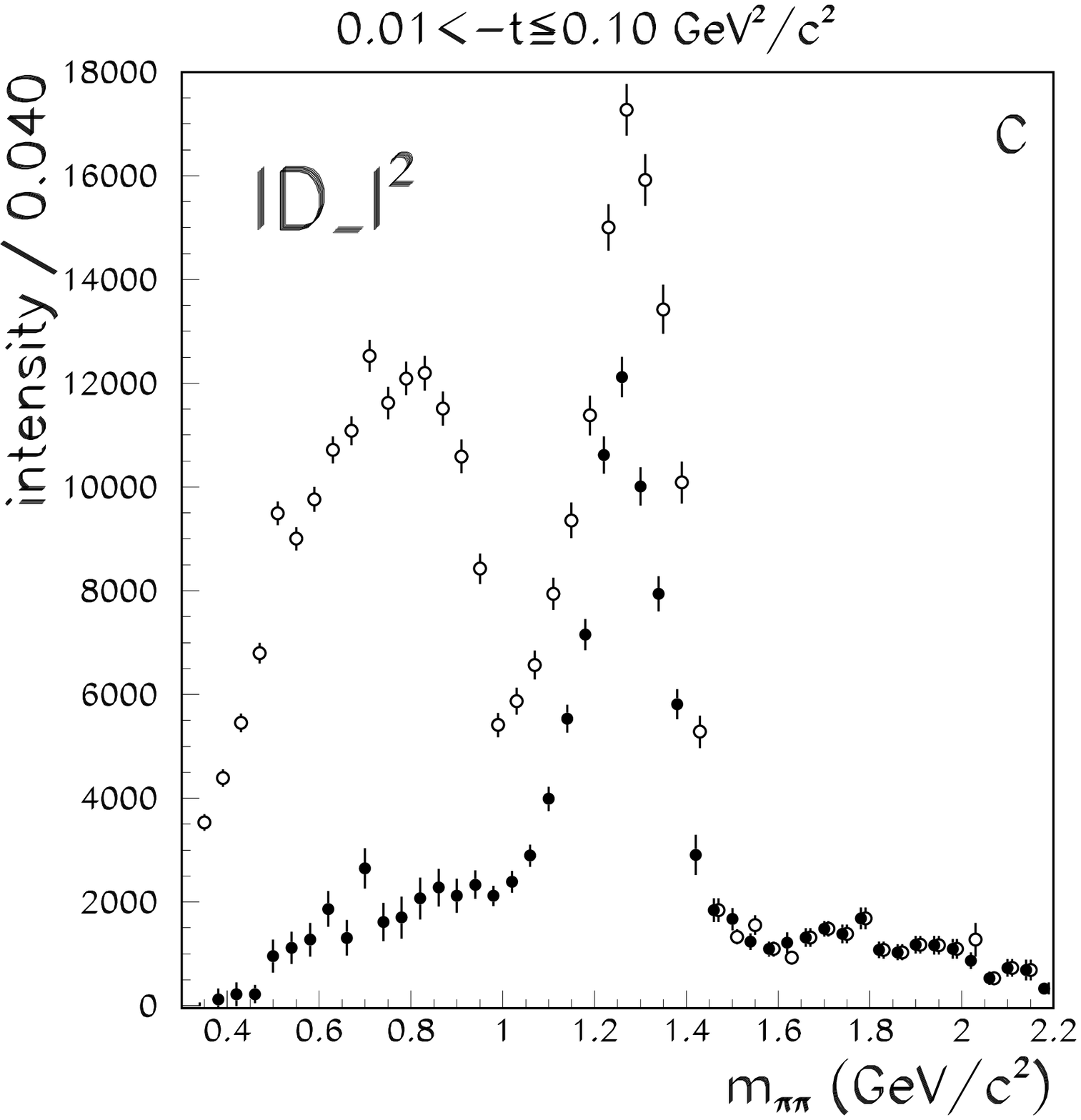,width=0.45\textwidth,height=0.45\textwidth}} &
\mbox{\epsfig{file=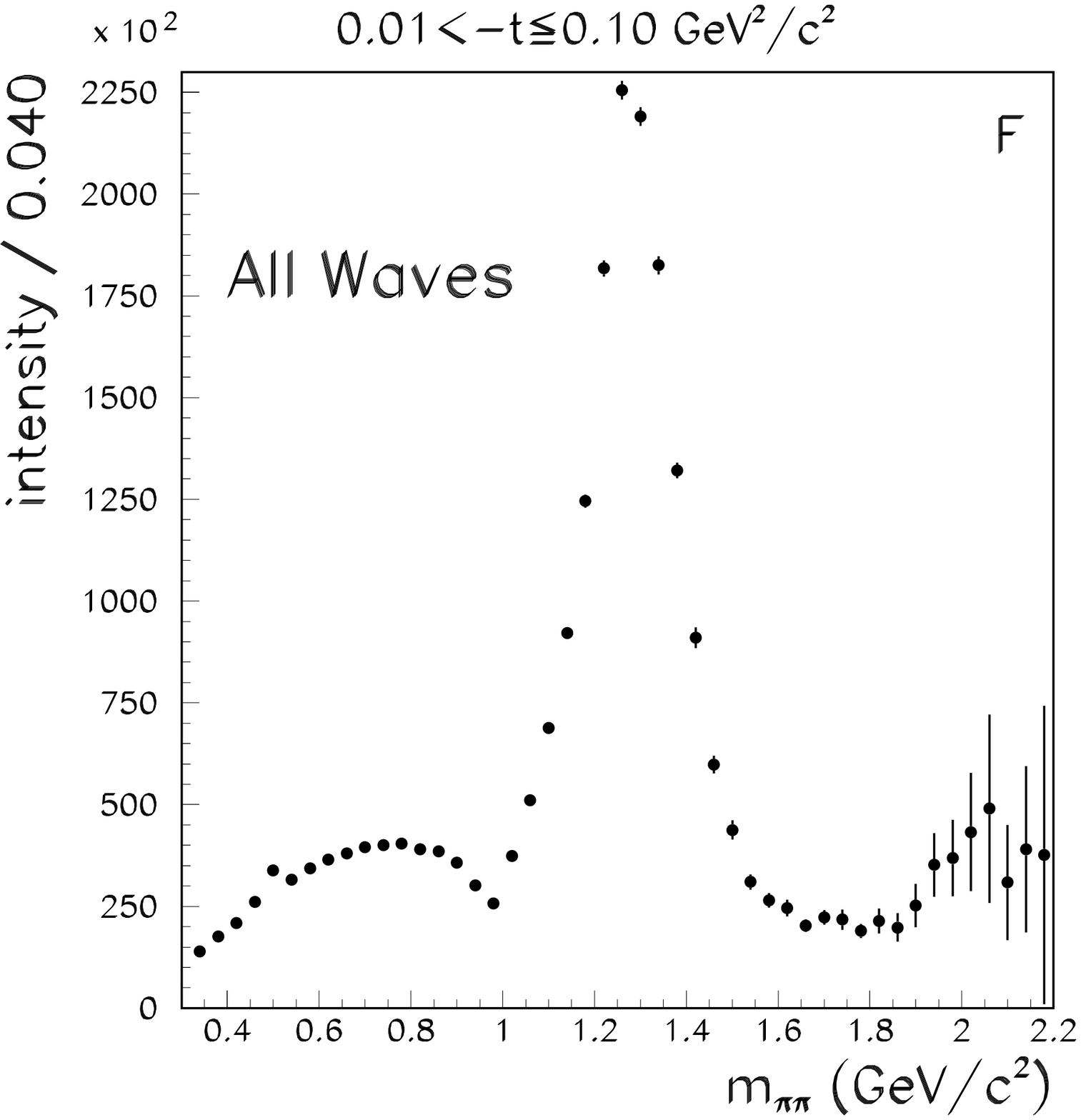,width=0.45\textwidth,height=0.45\textwidth}} 
\end{tabular}

\caption{\label{fig:t0intensities}The squares of the magnitudes of the 
partial waves (a)--(e) as a function of mass for events in the 
region \protect$ 0.01<|t|<0.10\, GeV^{2}/c^{2}\protect $.  
The solid circles correspond to the physical solution. 
The open circles correspond to the other ambiguous solution.
Additionally the coherent sum of the partial waves integrated over 
decay angles, (f),  gives the acceptance corrected mass distribution.  
The dominant production mechanism is \protect$ m=0\protect $ unnatural parity
exchange (\protect$ S\protect $ , \protect$ D_{0}\protect $ , and \protect$ G_{0}\protect $
partial waves). }

\end{figure}

\begin{figure}[htbp]\centering
\begin{tabular}{cc}
\mbox{\epsfig{file=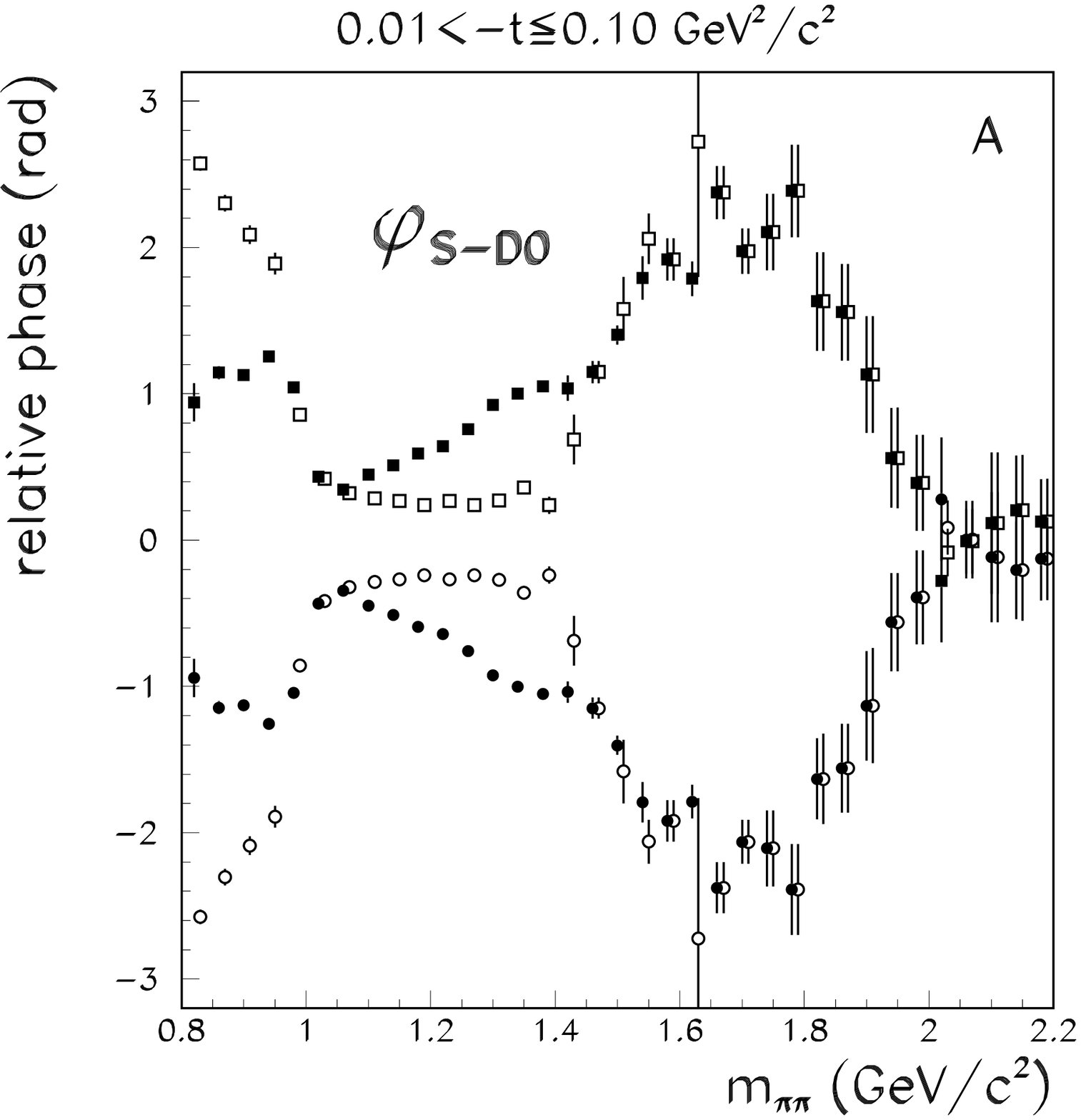,width=0.45\textwidth,height=0.45\textwidth}} &
\mbox{\epsfig{file=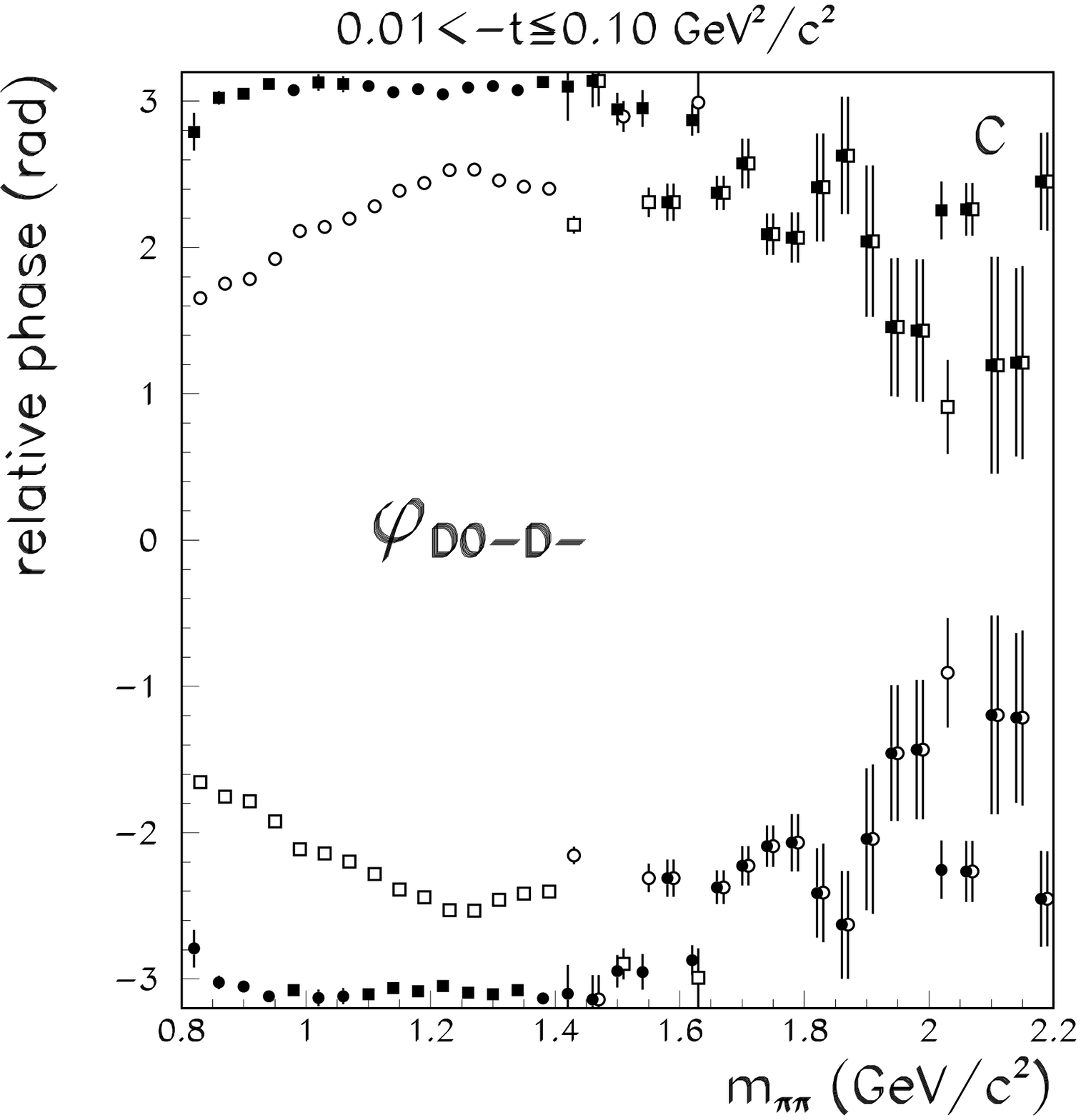,width=0.45\textwidth,height=0.45\textwidth}} \\
\mbox{\epsfig{file=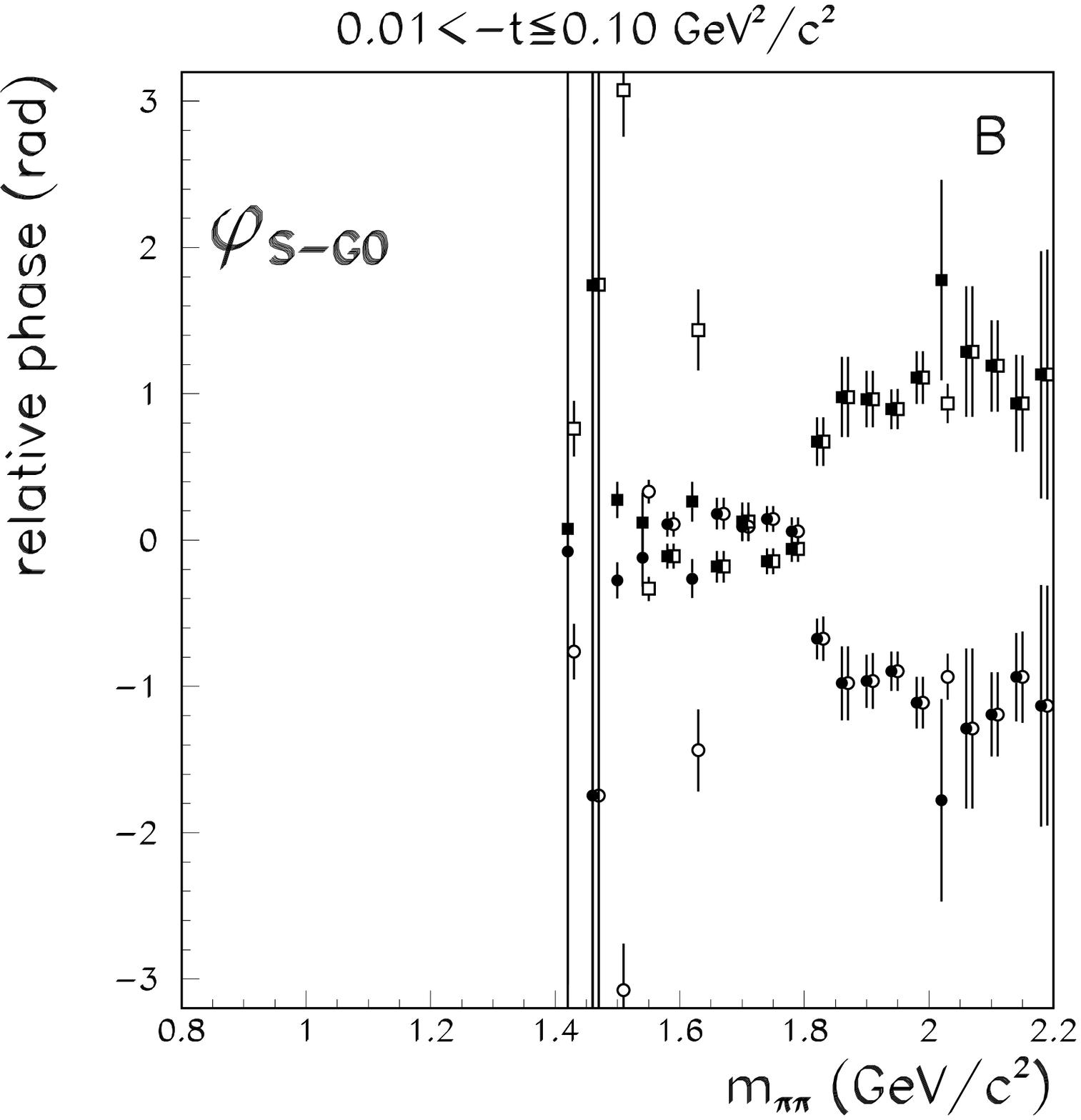,width=0.45\textwidth,height=0.45\textwidth}} \\
\end{tabular}

\caption{\label{fig:t0phases}For events in the region \protect$ 0.01<|t|<0.10\,$ 
the relative phase between the $S$ and $D_0$ partial waves (a) shows rapid phase variation 
near $0.98\,GeV/c^2$ and $1.5\,GeV/c^2$.  The relative phase between
the $D_0$ and $D_-$ partial waves (b) is smooth and nearly constant up to $1.5\,GeV/c^2$. 
The relative phase between the $S$ and $G_0$ partial waves is shown in
(c).  The solid circles represent the physical solution. }
\end{figure}

The results of the partial wave decomposition are shown in figures \ref{fig:t0intensities}
and \ref{fig:t0phases}. 
The partial wave intensities are shown in figure \ref{fig:t0intensities} and the phase
differences in figure \ref{fig:t0phases}.  The phase difference plots are shown
above $\pi^{0} \pi^{0}$ masses of 0.8 $GeV/c^2$.  Below that value, where one of
the waves is very small, phase difference information is unreliable.
As discussed in Section \ref{sec:ambigs}, there is
a two-fold ambiguity in the intensities. The threshold behavior ($S$-wave dominance) and 
the resonant behavior of the $f_2(1270)$ are used to select the physical solution.
 Furthermore, since 
the resonant structures of both the $ D_{0} $ and $ D_{-} $ partial
waves are due to the $ f_{2}(1270) $, the relative phase between the $ D_{0} $
and $ D_{-} $ partial waves should be constant
and near $\pm \pi$ radians, according to the phase convention of \cite{Chu97}.
These assumptions allow the physical solution at low mass to be connected with
the solutions at higher mass. Above approximately $ 1.5\, GeV/c^{2} $,
the solutions become degenerate. The spin-4 $ G_{0} $ partial wave
is not included in the fit below  $ 1.4\, GeV/c^{2} $. 

There are a number of  key features observed in the physical solution.  
There is at least one 
broad enhancement in the $ S $-wave intensity and a sharp dip 
in the $ S $-wave intensity near $ 1.0\, GeV/c^{2} $
accompanied by rapid phase variation in the $ S- $ $ D_{0} $ relative
phase.
There  also exists  a dip in the $ S $-wave intensity near $ 1.5\, GeV/c^{2} $
accompanied by rapid phase variation in the $ S- $ $ D_{0} $ relative
phase.
      The $ f_{2}(1270) $ is observed in the $ D_{0},D_{-} $, and $ D_{+} $
partial wave intensities, and 
the  bump  observed in the $ G_{0} $ partial wave
near 2.0 $GeV/c^2$ is consistent with the $ f_{4}(2040) $.  Finally, the
$D_0$-wave intensity is larger than the $D_{-}$-wave intensity or the 
$D_+$-wave intensity, consistent with the expectation
that OPE should favor production of an m=0 wave for this low-$|t|$
region.

A background term was not included in the PWA fits presented in this
paper.  A background term was included in some earlier fits where it
was found that below about 1.0 $GeV/c^2$ it cannot be distinguished from
the dominant $S_{0}$ wave and above 1.0 $GeV/c^2$, the fit forces the 
background term to zero.

\subsubsection{Results for \protect$ 0.10<-t<0.20\, GeV^{2}/c^{2}\protect $ }

\begin{figure}[htbp]\centering
\begin{tabular}{cc}
\mbox{\epsfig{file=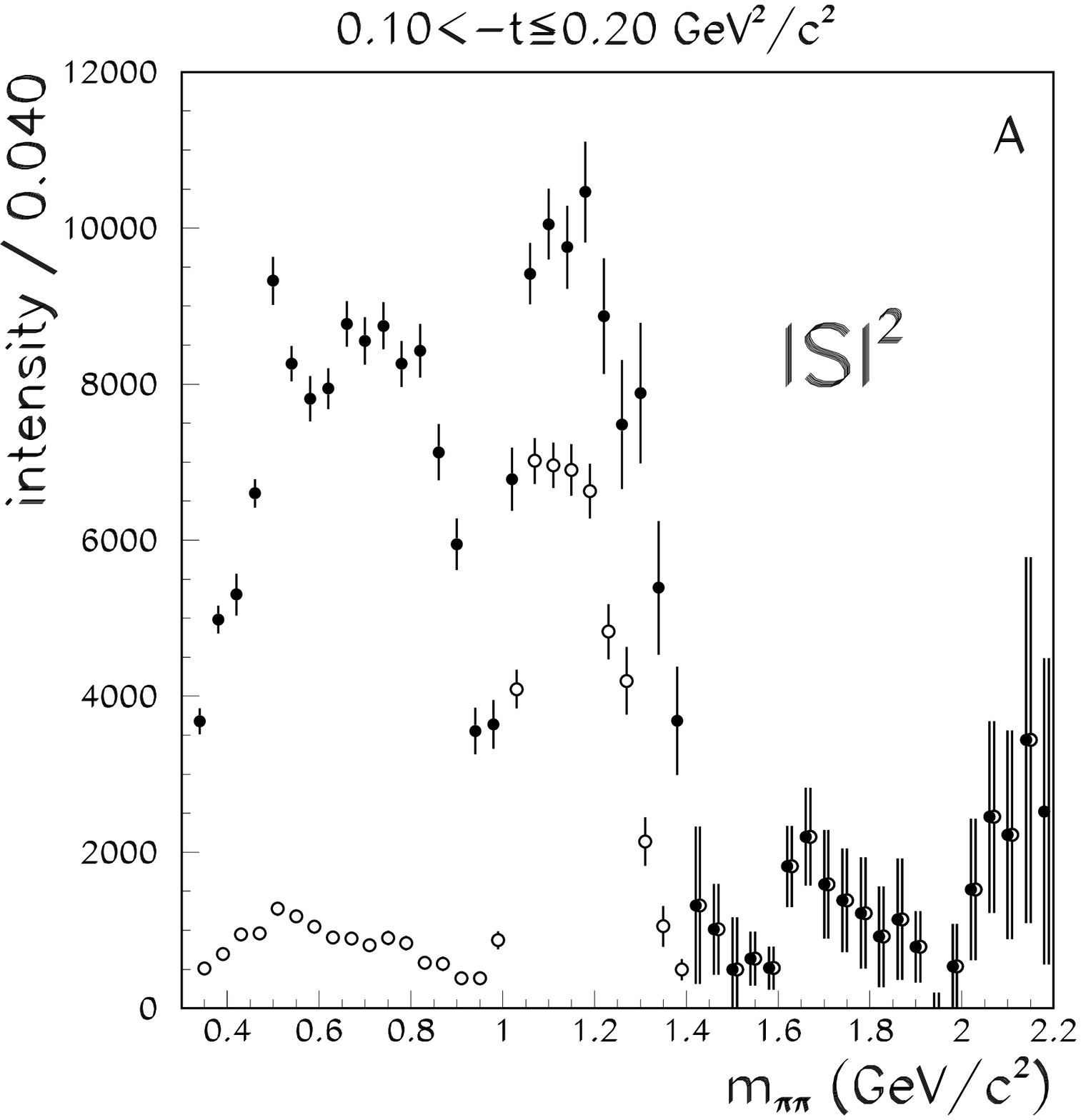,width=0.45\textwidth,height=0.45\textwidth}} &
\mbox{\epsfig{file=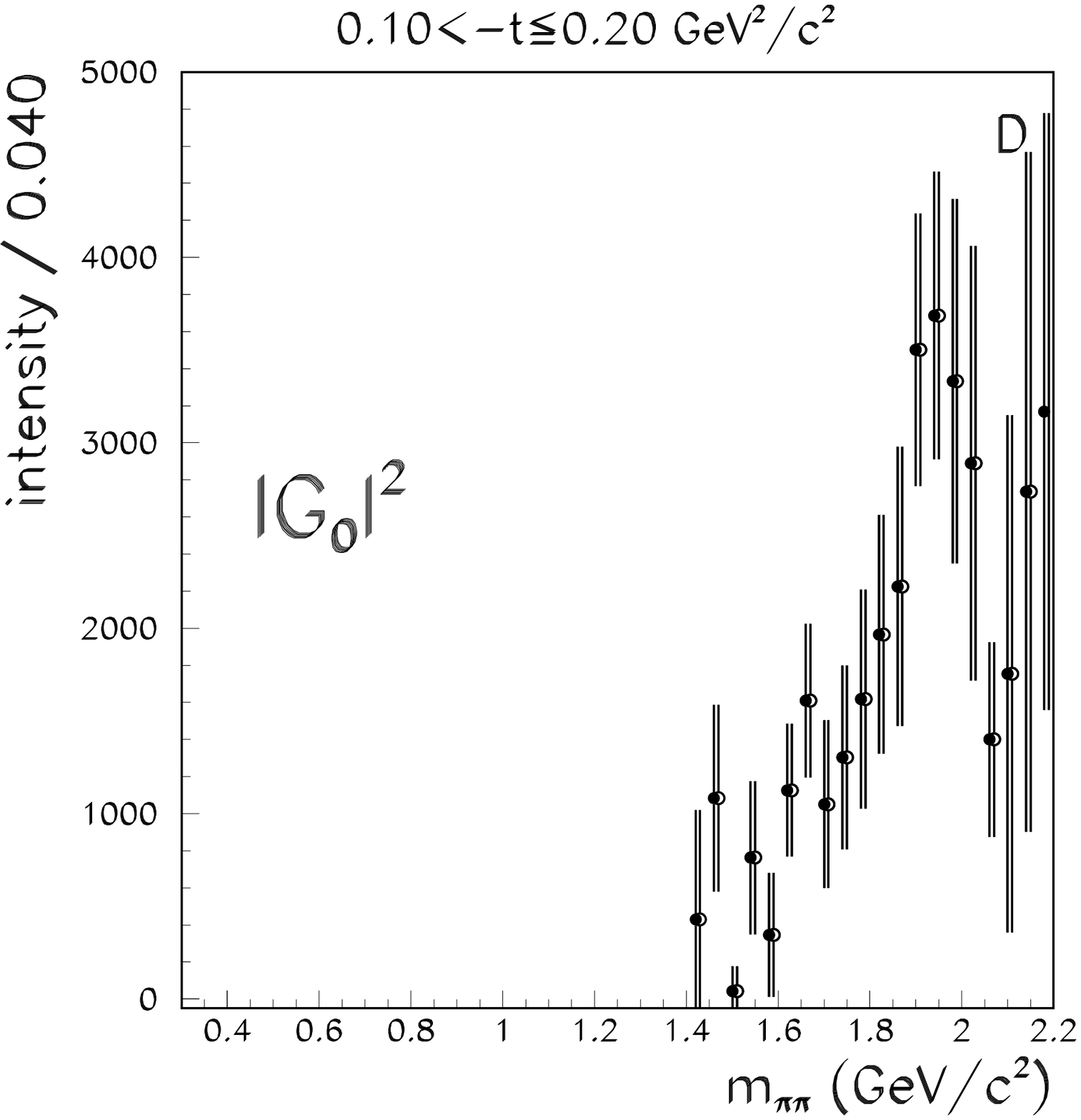,width=0.45\textwidth,height=0.45\textwidth}} \\
\mbox{\epsfig{file=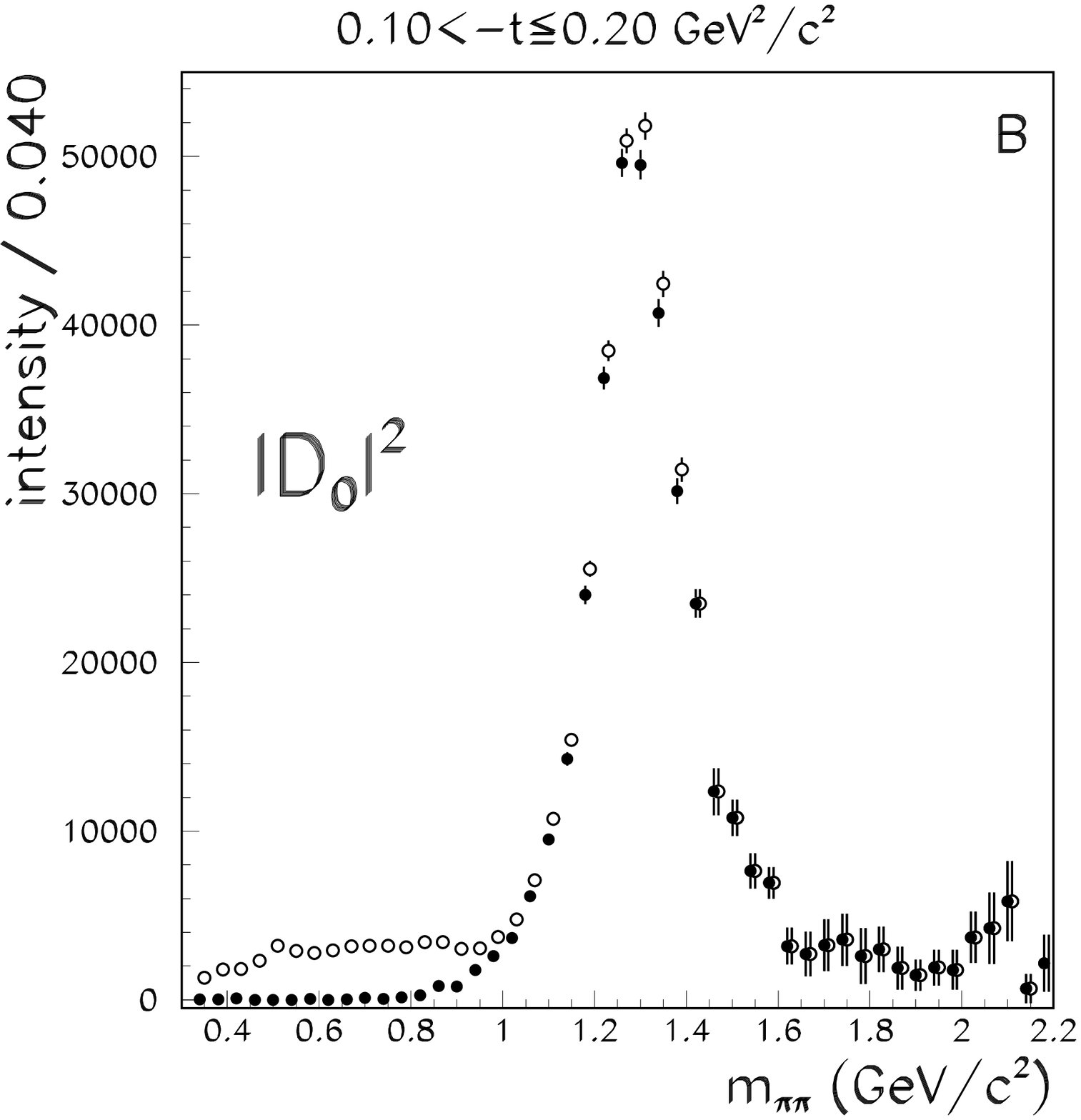,width=0.45\textwidth,height=0.45\textwidth}} &
\mbox{\epsfig{file=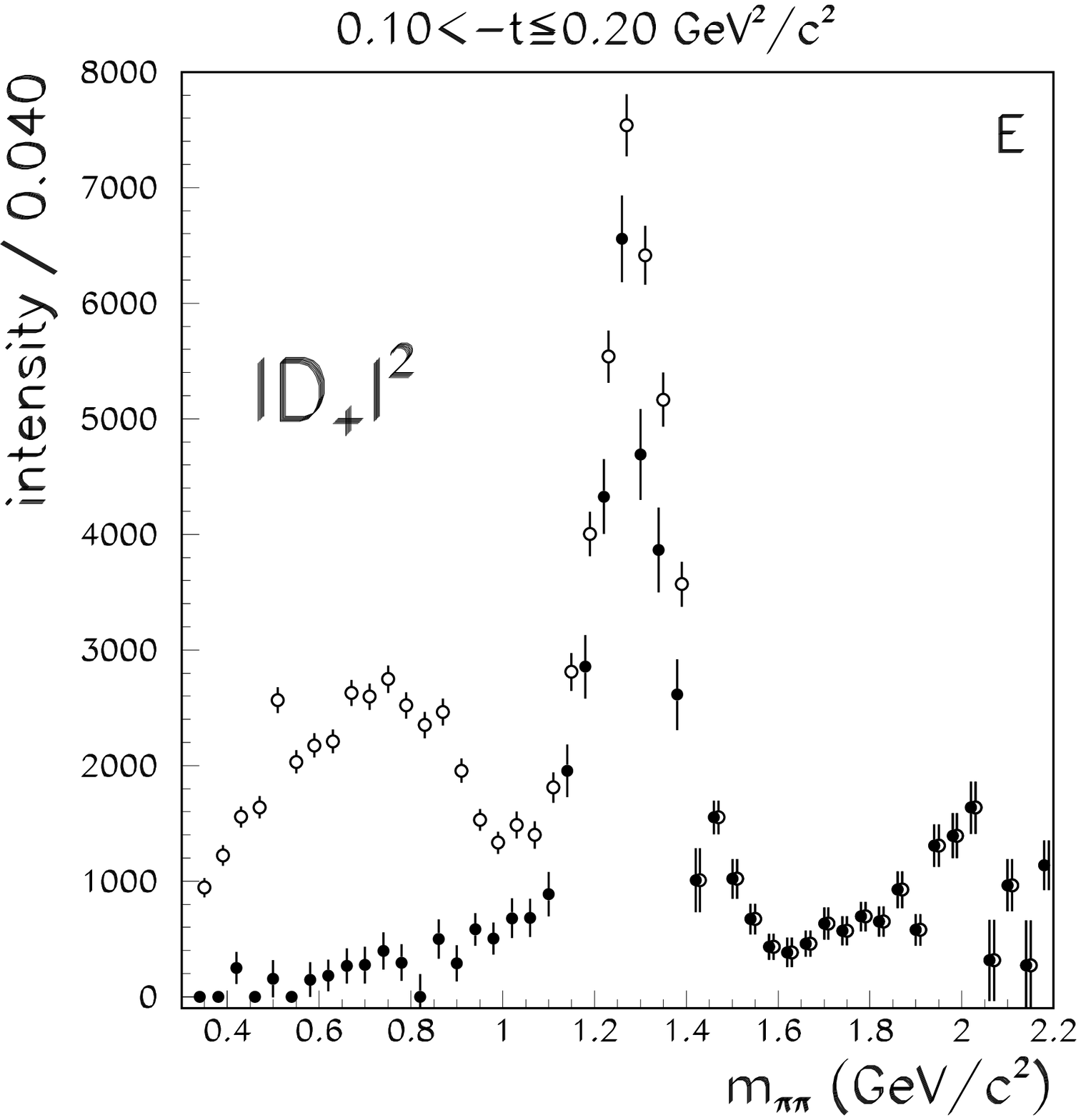,width=0.45\textwidth,height=0.45\textwidth}} \\ 
\mbox{\epsfig{file=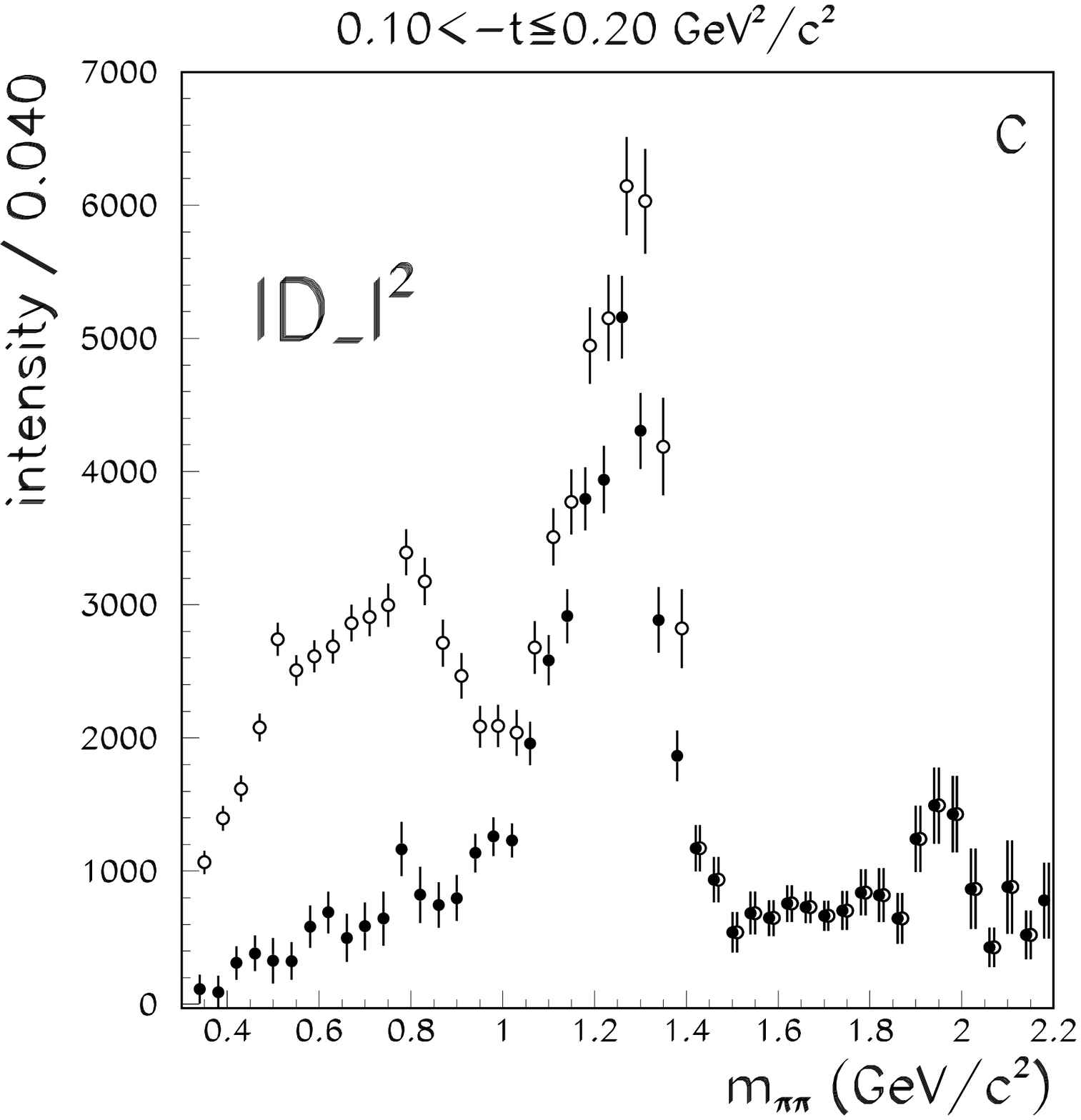,width=0.45\textwidth,height=0.45\textwidth}} &
\mbox{\epsfig{file=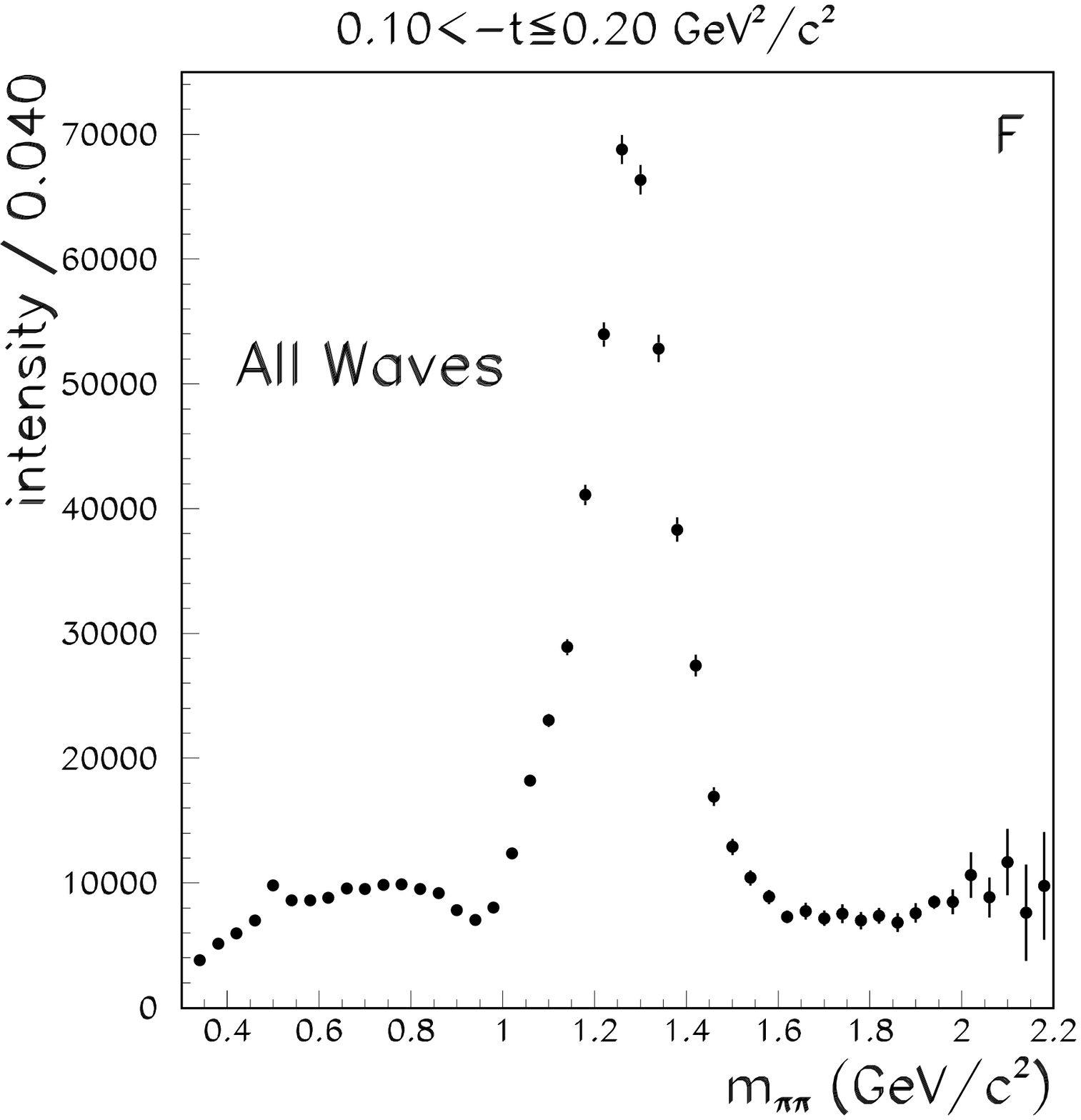,width=0.45\textwidth,height=0.45\textwidth}} 
\end{tabular}

\caption{\label{fig:t1intensities}The squares of the magnitudes of the partial 
waves  (a)--(e) as a function of mass for events in the region \protect$
0.10<|t|<0.20\,  GeV^{2}/c^{2}\protect $.  The solid circles correspond
to the physical solution. The coherent sum of the partial waves
integrated over decay angles, (f),  gives the acceptance corrected mass
distribution.    As in figure \ref{fig:t0intensities}, the dominant
production mechanism  is \protect$ m=0\protect $ unnatural parity
exchange (\protect$ S\protect $ , \protect$ D_{0}\protect $ , and
\protect$ G_{0}\protect $ partial waves). }

\end{figure}

\begin{figure}[htbp]\centering
\begin{tabular}{cc}
\mbox{\epsfig{file=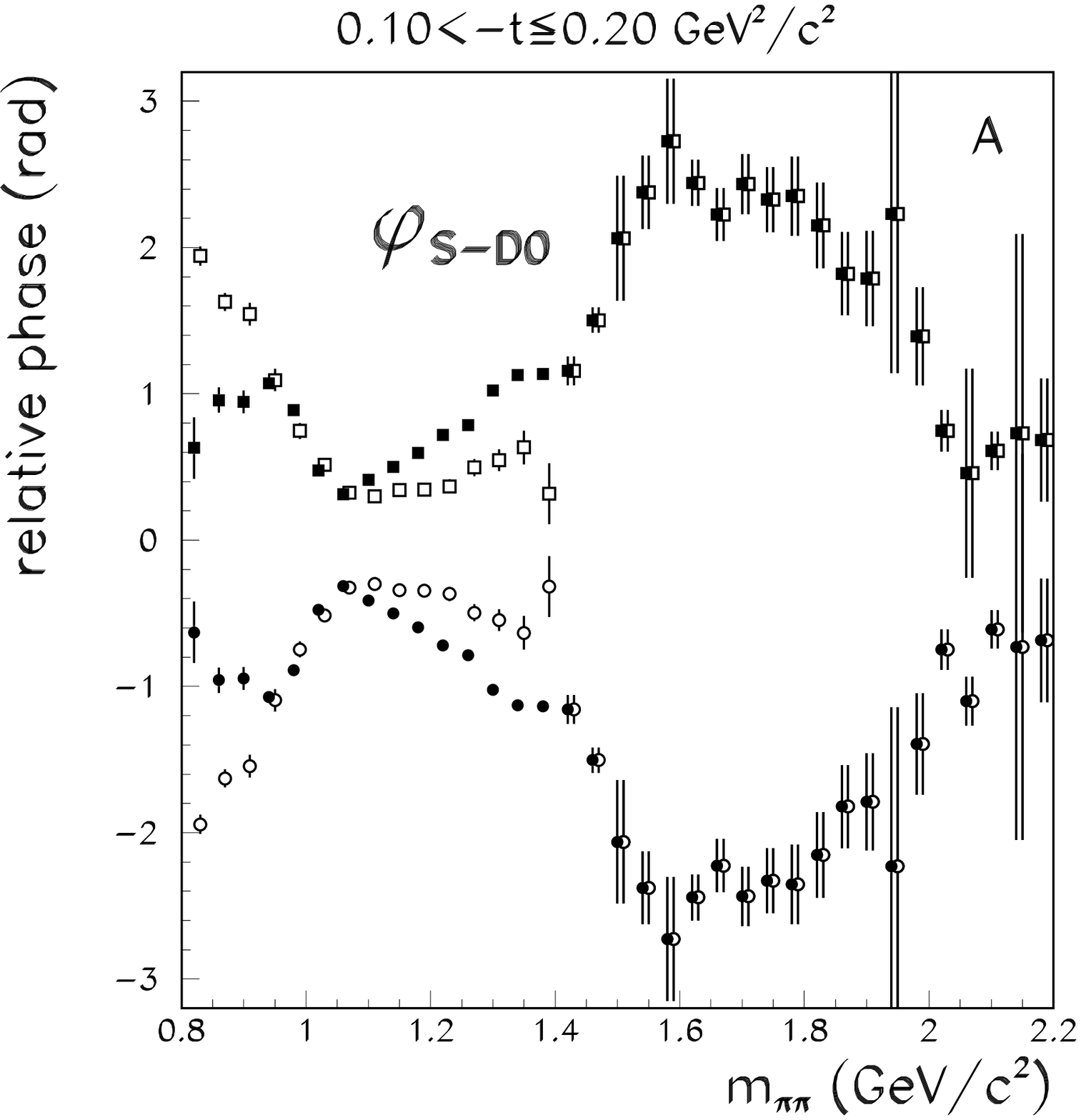,width=0.45\textwidth,height=0.45\textwidth}} &
\mbox{\epsfig{file=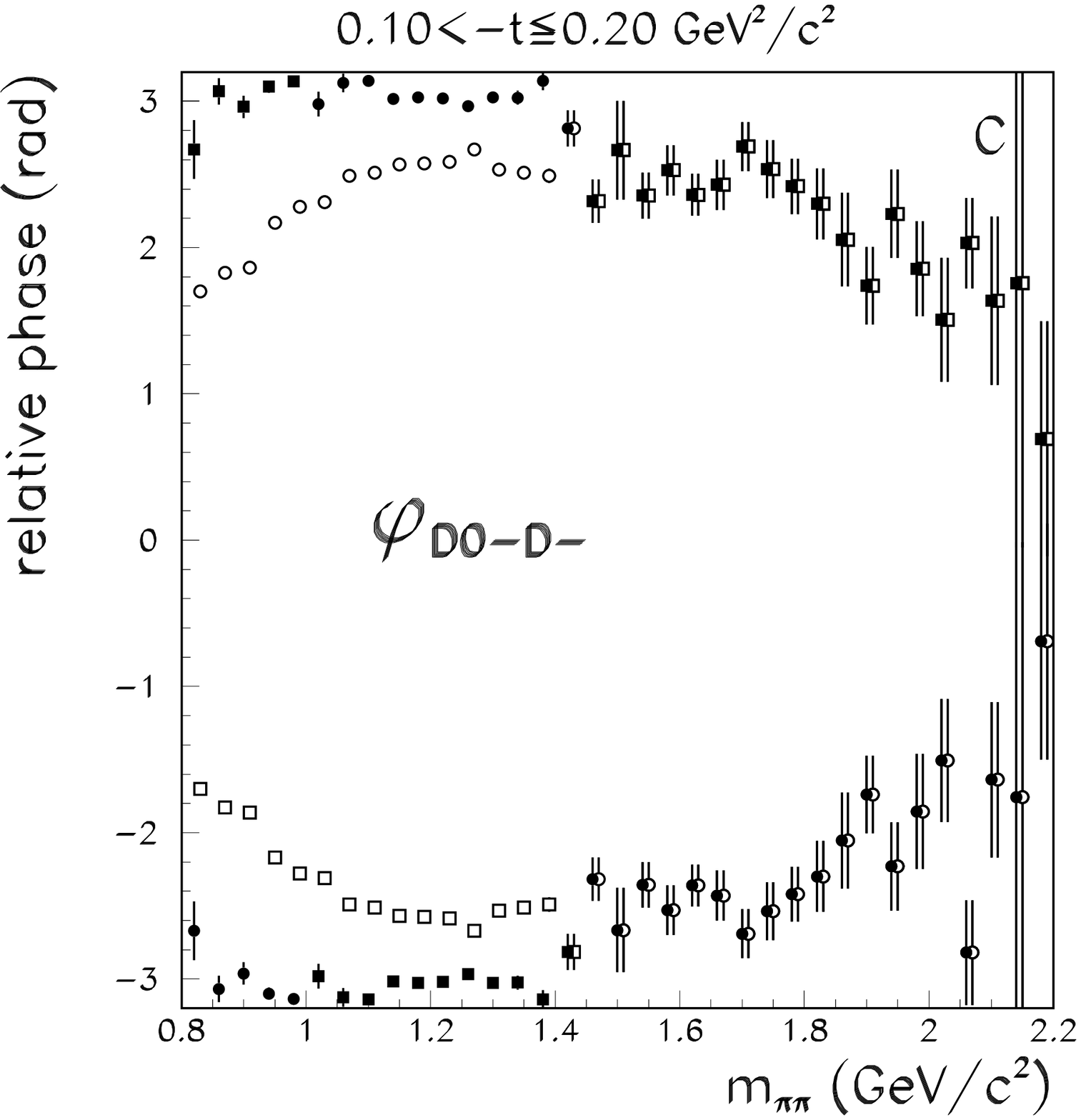,width=0.45\textwidth,height=0.45\textwidth}} \\
\mbox{\epsfig{file=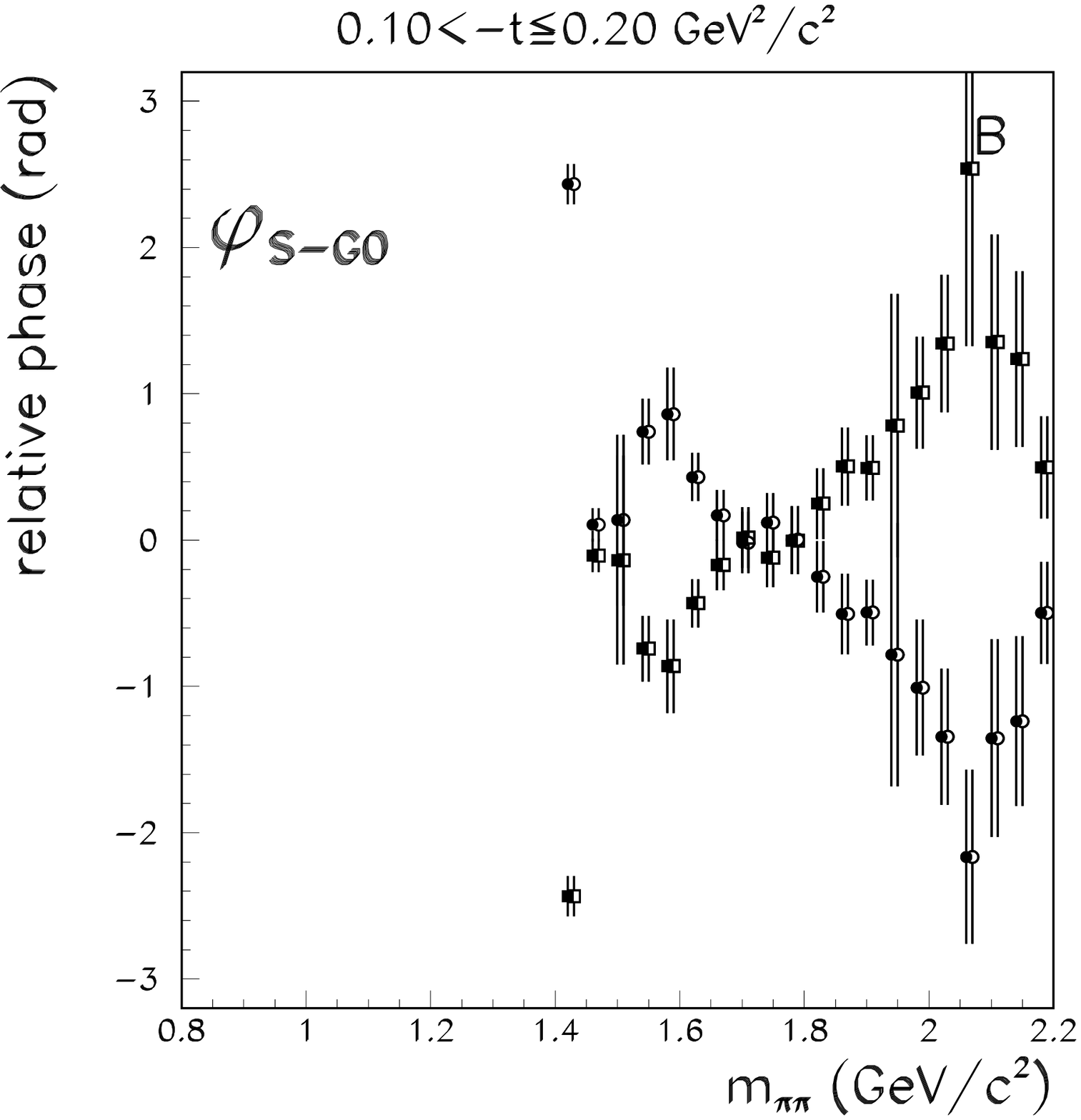,width=0.45\textwidth,height=0.45\textwidth}} \\
\end{tabular}

\caption{\label{fig:t1phases}For events in the region \protect$ 0.10<|t|<0.20\,$ the 
relative phase between the $S$ and $D_0$ partial waves (a) shows rapid phase variation 
near $0.98\,GeV/c^2$ and $1.5\,GeV/c^2$.  The relative phase between the
$D_0$ and $D_-$ partial waves (b) is smooth and nearly constant up to $1.5\,GeV/c^2$.  
The relative phase between the $S$ and $G_0$ partial waves is shown in
(c).  The solid circles represent the physical solution. }

\end{figure}

The results of the partial wave analysis (Figures \ref{fig:t1intensities} and
\ref{fig:t1phases}) in this region are qualitatively similar to the results
in the $ 0.01<-t<0.10\, GeV^{2}/c^{2} $ region. The same techniques are used to select 
the physical solution as in the previous region in $|t|$. The $ S $-wave intensity
 contains at least one broad object and two dips. The $ f_{2}(1270) $
is  observed in all $ D $-waves. An enhancement near
2.0 $GeV/c^2$  is again observed
in the $ G_{0} $ partial wave. More detailed comparisons with the results
from the $ 0.01<-t<0.10\, GeV/c^{2} $ region reveal the following differences: The ratio 
of the $ S $-wave intensity to the $ D_{0} $-wave intensity is smaller
at larger $ |t| $ and 
the ratio of the $ D_{0} $-wave intensity to both the 
$ D_{-} $-wave and  $ D_{+} $-wave intensities is 
smaller at
larger $ |t| $. 
The ratio of the $ D_{0} $-wave intensity to 
$ G_{0} $-wave intensity does not change suggesting 
that the $ f_{2}(1270) $ and the $ f_{4}(2040) $ are produced
by the same mechanism. 

\subsubsection{Results for \protect$ 0.20<-t<0.40\, GeV^{2}/c^{2}\protect $ }

\begin{figure}[htbp]\centering
\begin{tabular}{cc}
\mbox{\epsfig{file=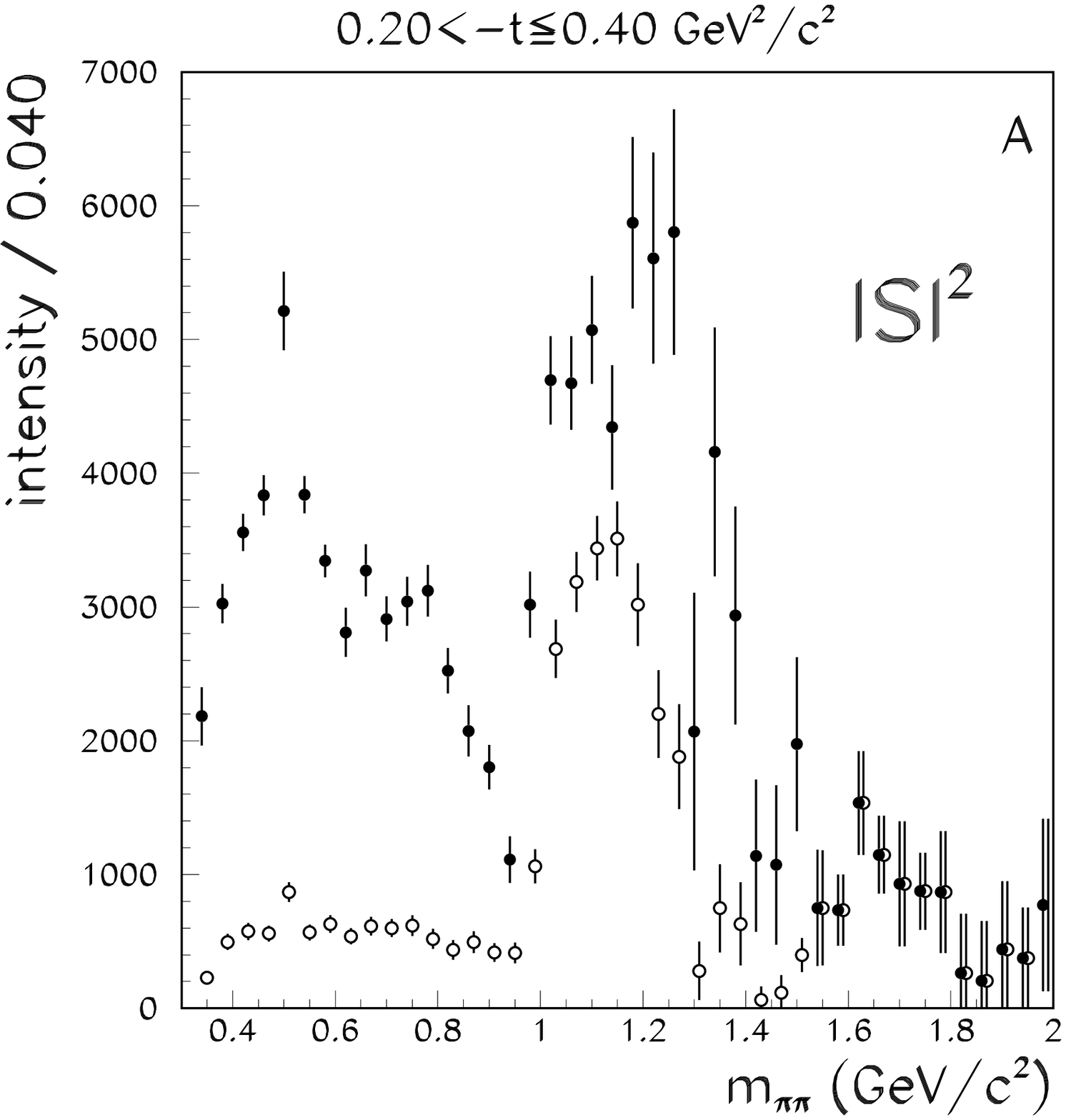,width=0.45\textwidth,height=0.45\textwidth}} \\
\mbox{\epsfig{file=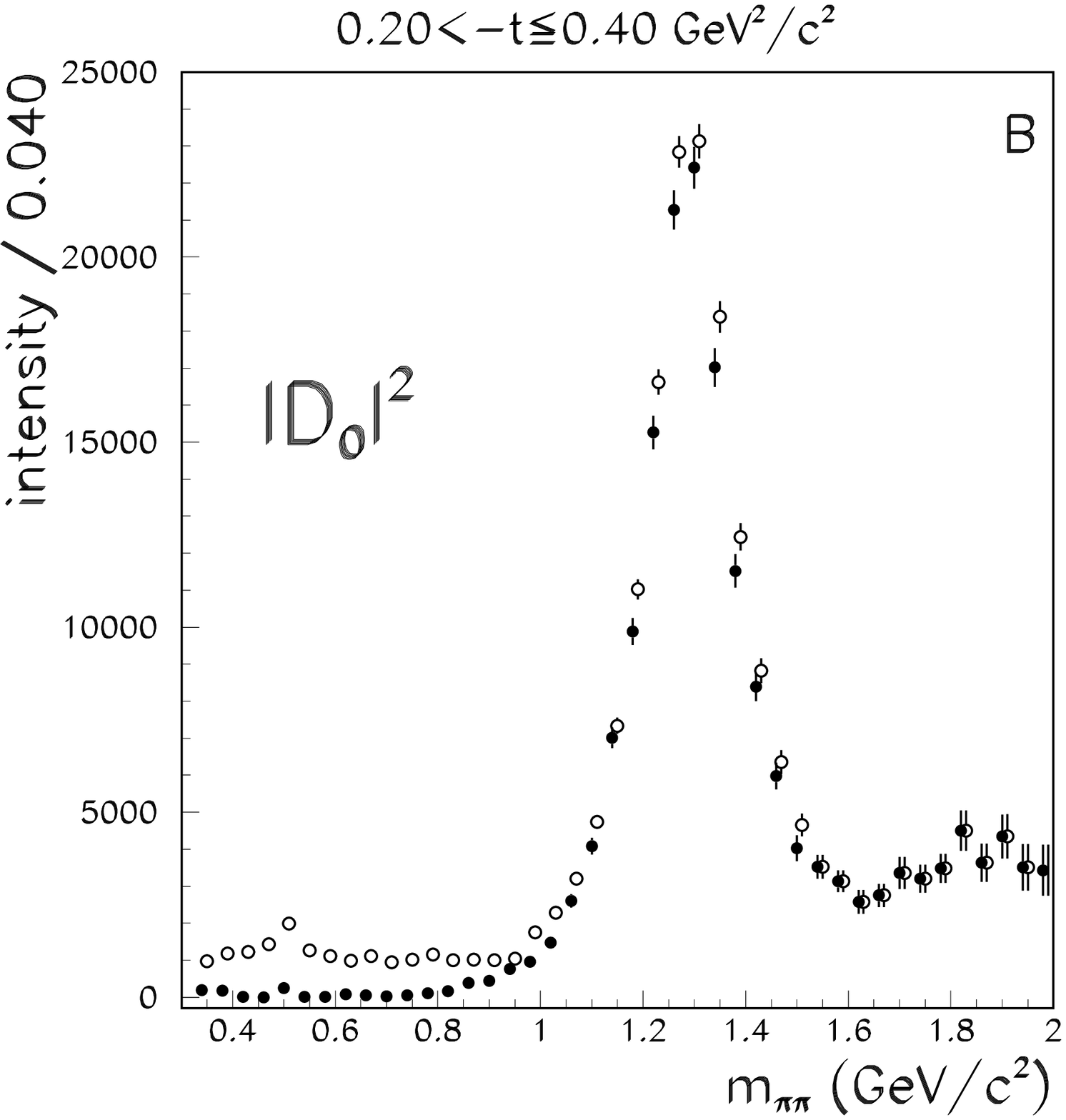,width=0.45\textwidth,height=0.45\textwidth}} &
\mbox{\epsfig{file=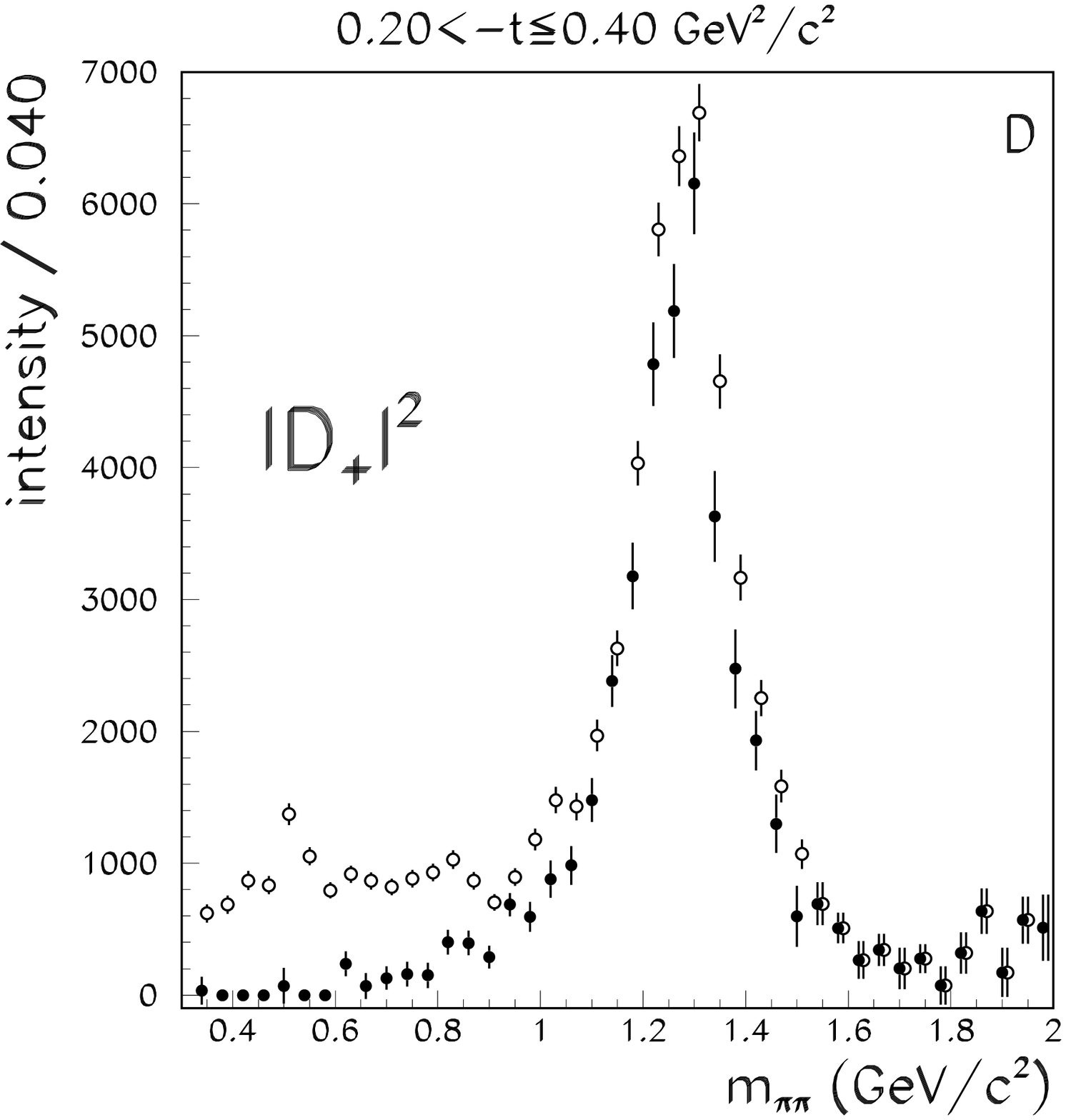,width=0.45\textwidth,height=0.45\textwidth}} \\ 
\mbox{\epsfig{file=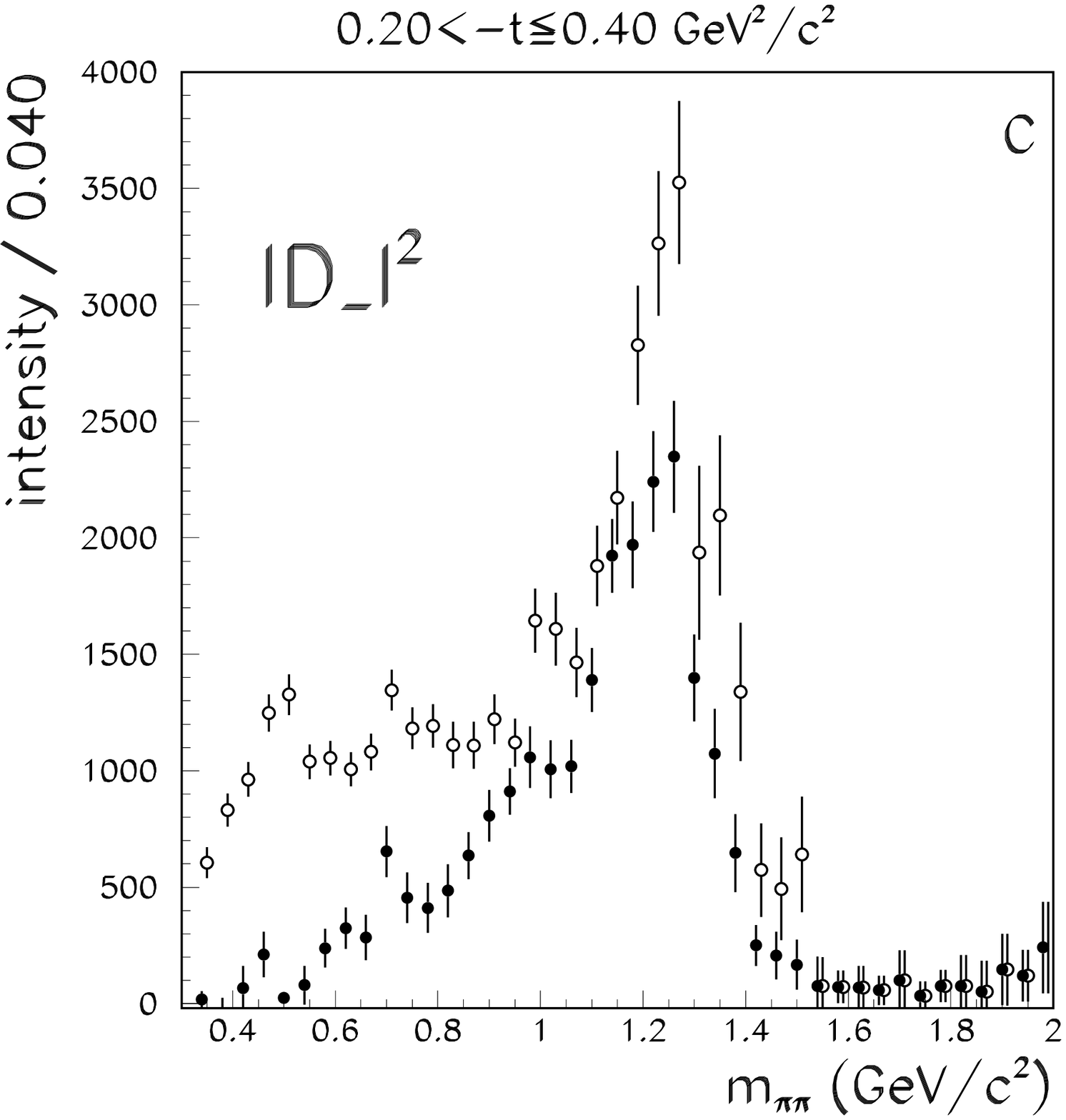,width=0.45\textwidth,height=0.45\textwidth}} &
\mbox{\epsfig{file=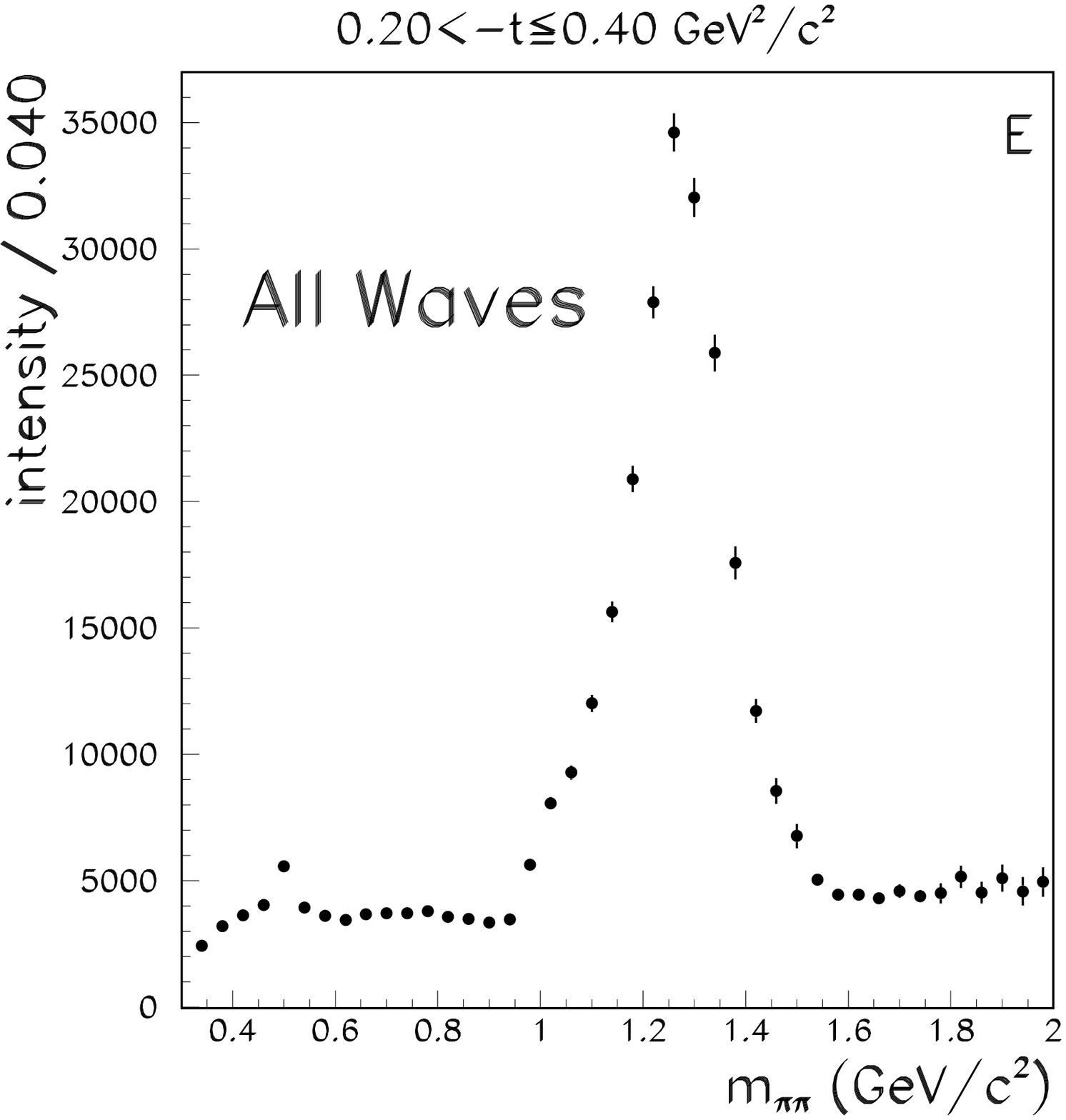,width=0.45\textwidth,height=0.45\textwidth}} 
\end{tabular}

\caption{\label{fig:t2intensities}The squares of the magnitudes of the 
partial waves  (a)--(d) as a function of mass for events in the
 region \protect$ 0.20<|t|<0.40\, GeV^{2}/c^{2}\protect $.  
The solid circles correspond to the physical solution. The coherent sum
of the  partial waves integrated
over decay angles, (e),  gives the acceptance corrected mass distribution.   
Compared to figures \ref{fig:t0intensities} and \ref{fig:t1intensities} the $D_+$
partial wave (natural parity exchange) is becoming more important although 
the dominant production mechanism is  still \protect$ m=0\protect $ unnatural parity
exchange (\protect$ S\protect $ , \protect$ D_{0}\protect $ , and \protect$ G_{0}\protect $ partial waves). }

\end{figure}

\begin{figure}[htbp]\centering
\begin{tabular}{cc}
\mbox{\epsfig{file=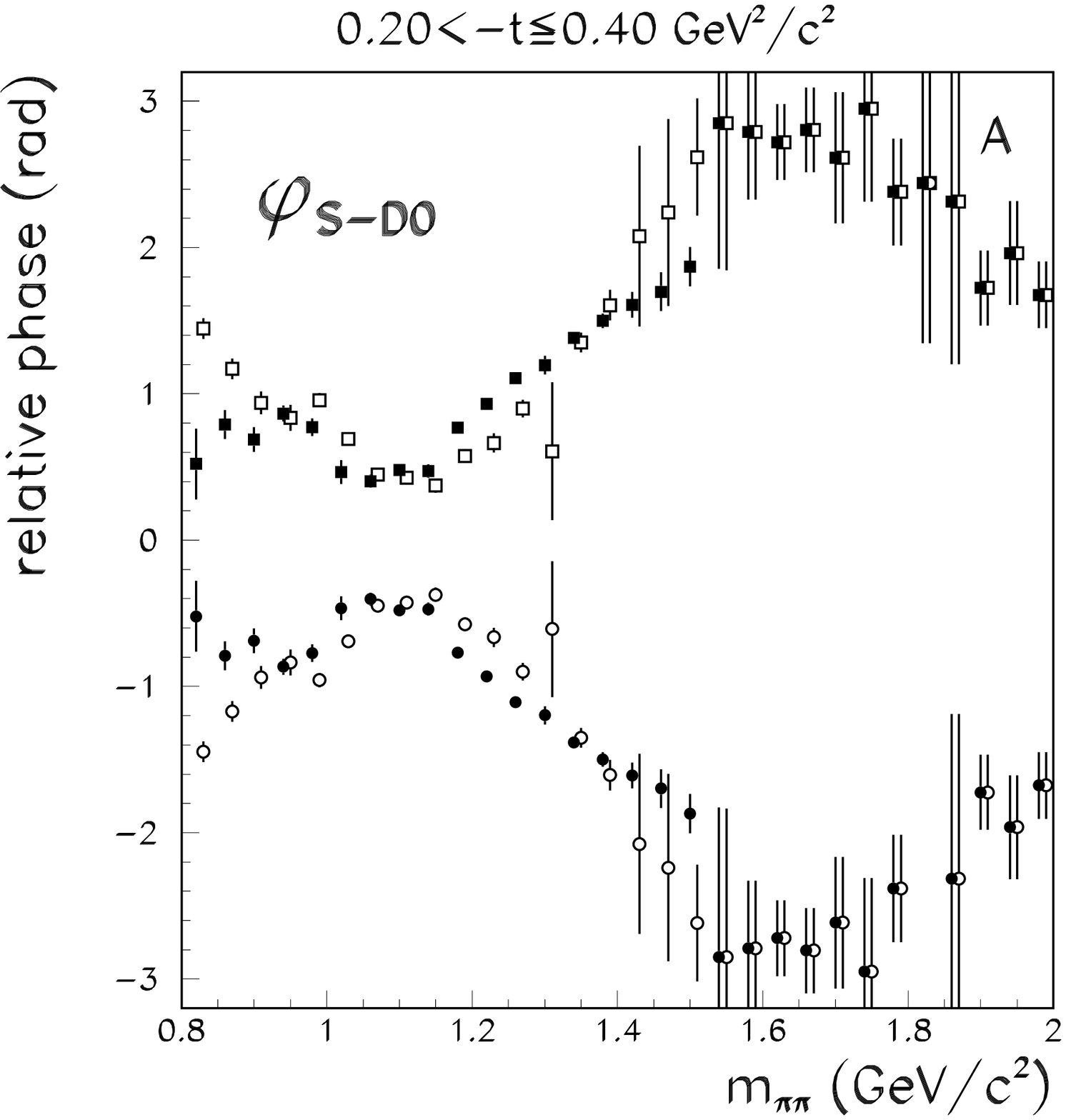,width=0.45\textwidth,height=0.45\textwidth}} &
\mbox{\epsfig{file=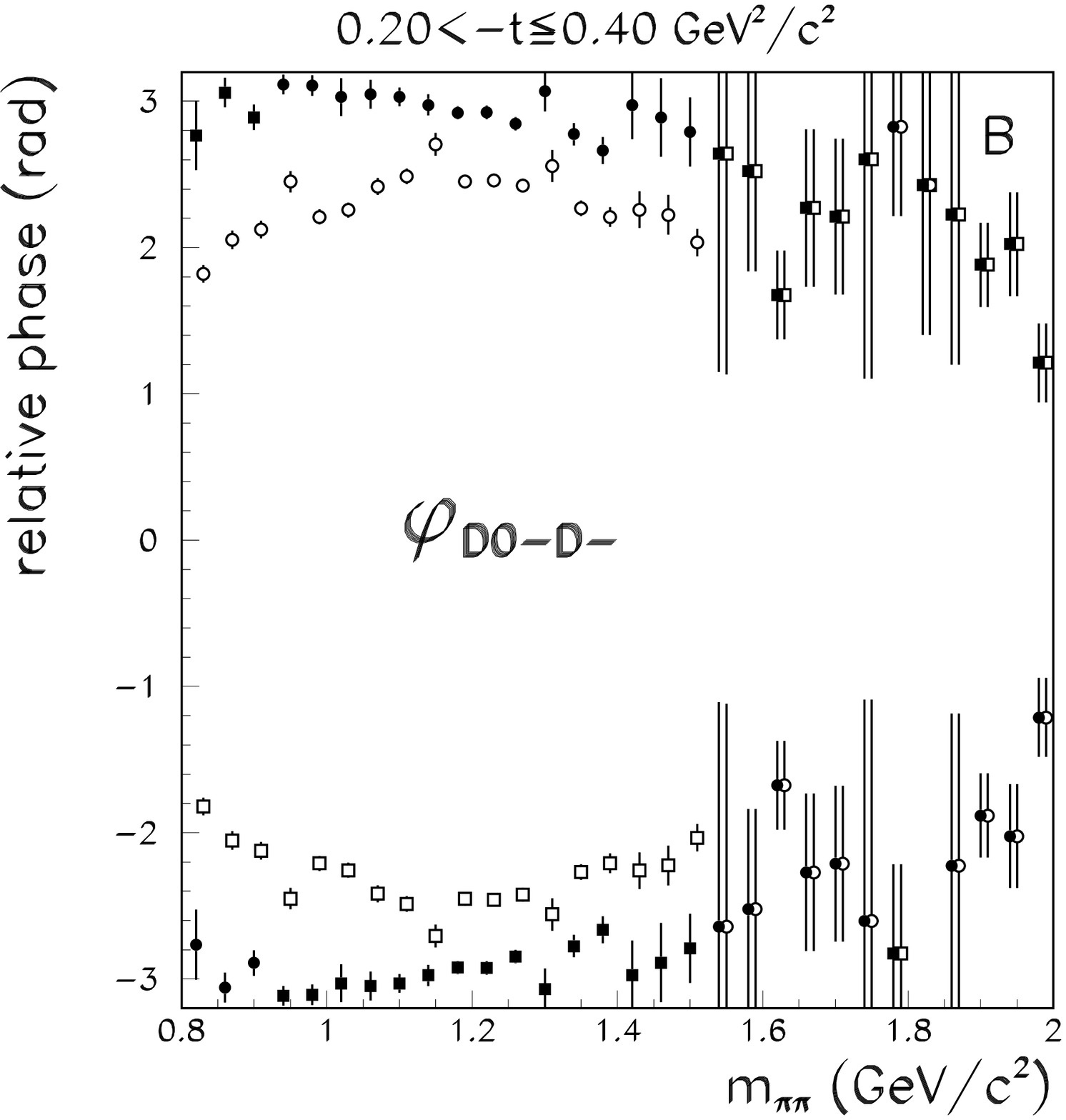,width=0.45\textwidth,height=0.45\textwidth}} \\
\end{tabular}

\caption{\label{fig:t2phases}The relative phases between unnatural parity exchange
partial waves  for events in the region \protect$ 0.20<|t|<0.40$ .  
The physical solution (solid circles) in the $S-D_0$ relative phase plot (a) 
shows less rapid phase variation than in figures \ref{fig:t0phases} and
\ref{fig:t1phases}. The $D_0-D_-$ relative phase is shown in (b).}

\end{figure}

The change in slope for the $| t| $-distribution as seen in figure \ref{fig:m&t}, indicates a 
change
in production mechanism. This is reflected in the partial wave analysis as well (Figures 
\ref{fig:t2intensities}
and \ref{fig:t2phases}). 
For this $|t|$ region, the $G_0$ partial wave is not required for an adequate
description of the observed angular distributions and is therefore not included
here or in the next higher $|t|$ region.
 The $ S $-wave intensity has a 
different shape compared to that at smaller values
of $|t| $. The $ D_{+} $-wave intensity is approximately one-third as large
as the $ D_{0} $-wave intensity whereas at smaller momentum transfer it was approximately
one-tenth as large.

\subsubsection{Results for \protect$ 0.40<-t<1.50\, GeV^{2}/c^{2}\protect $ }

\begin{figure}[htbp]\centering
\begin{tabular}{cc}
\mbox{\epsfig{file=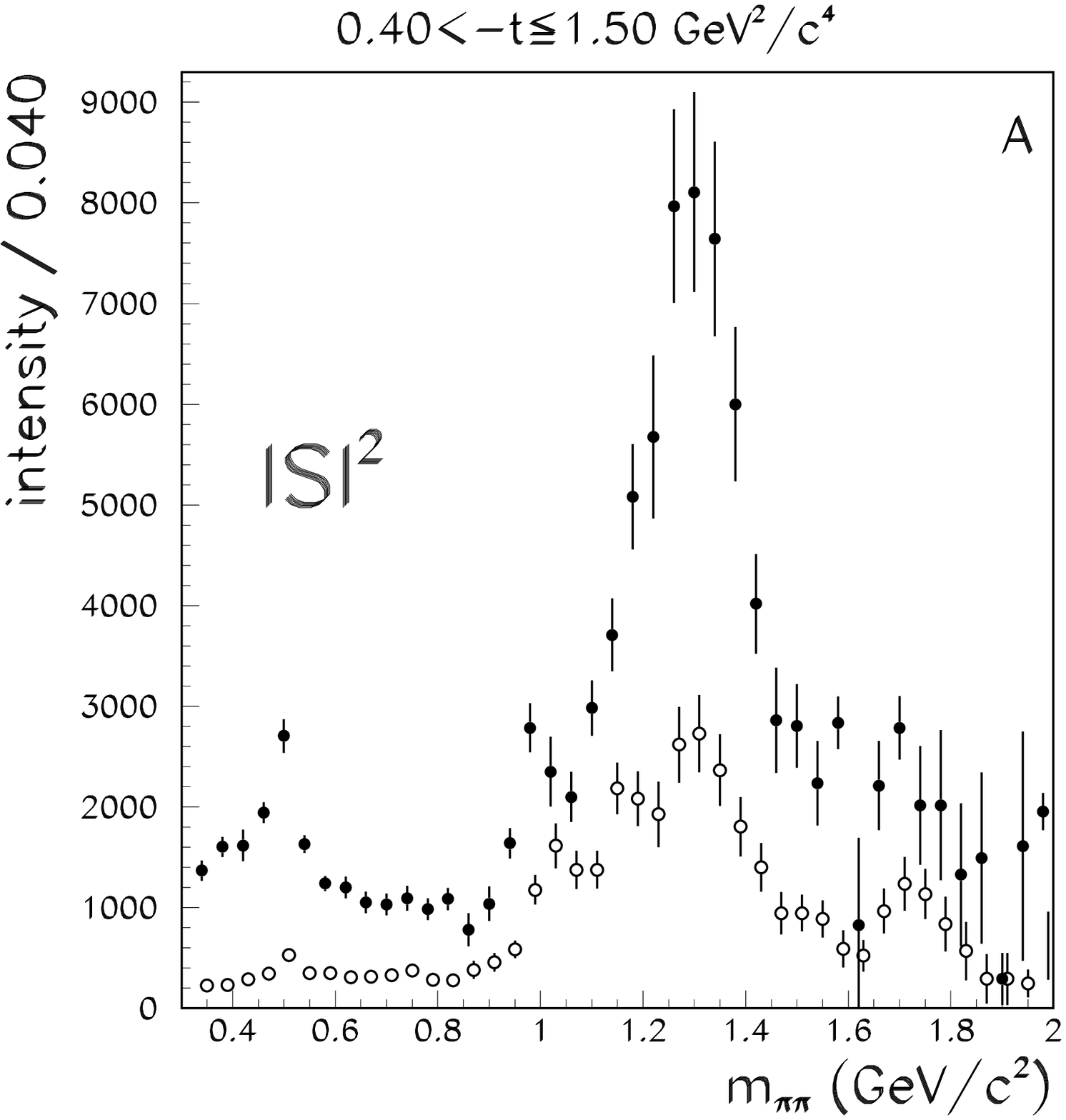,width=0.45\textwidth,height=0.45\textwidth}} \\
\mbox{\epsfig{file=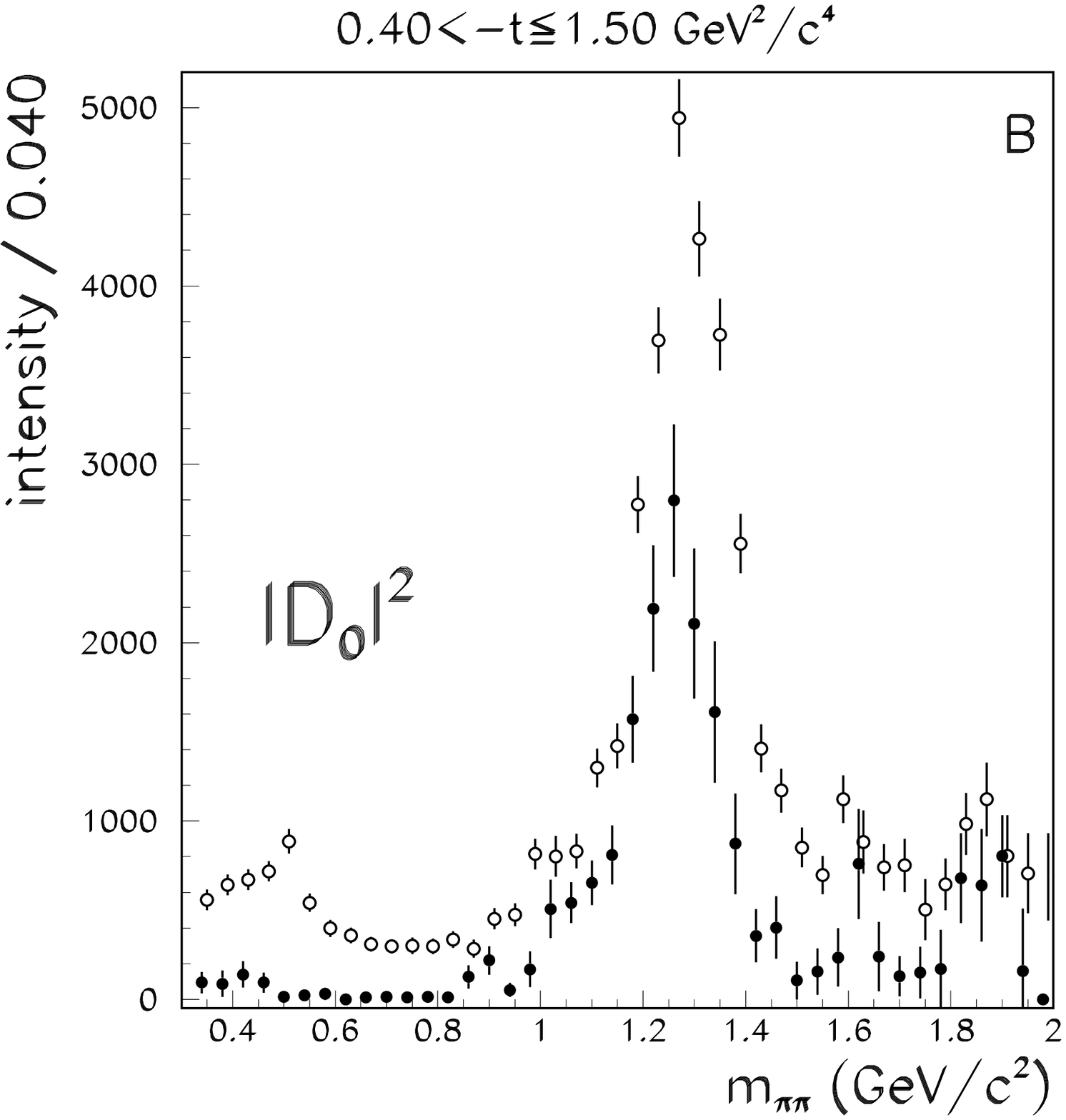,width=0.45\textwidth,height=0.45\textwidth}} &
\mbox{\epsfig{file=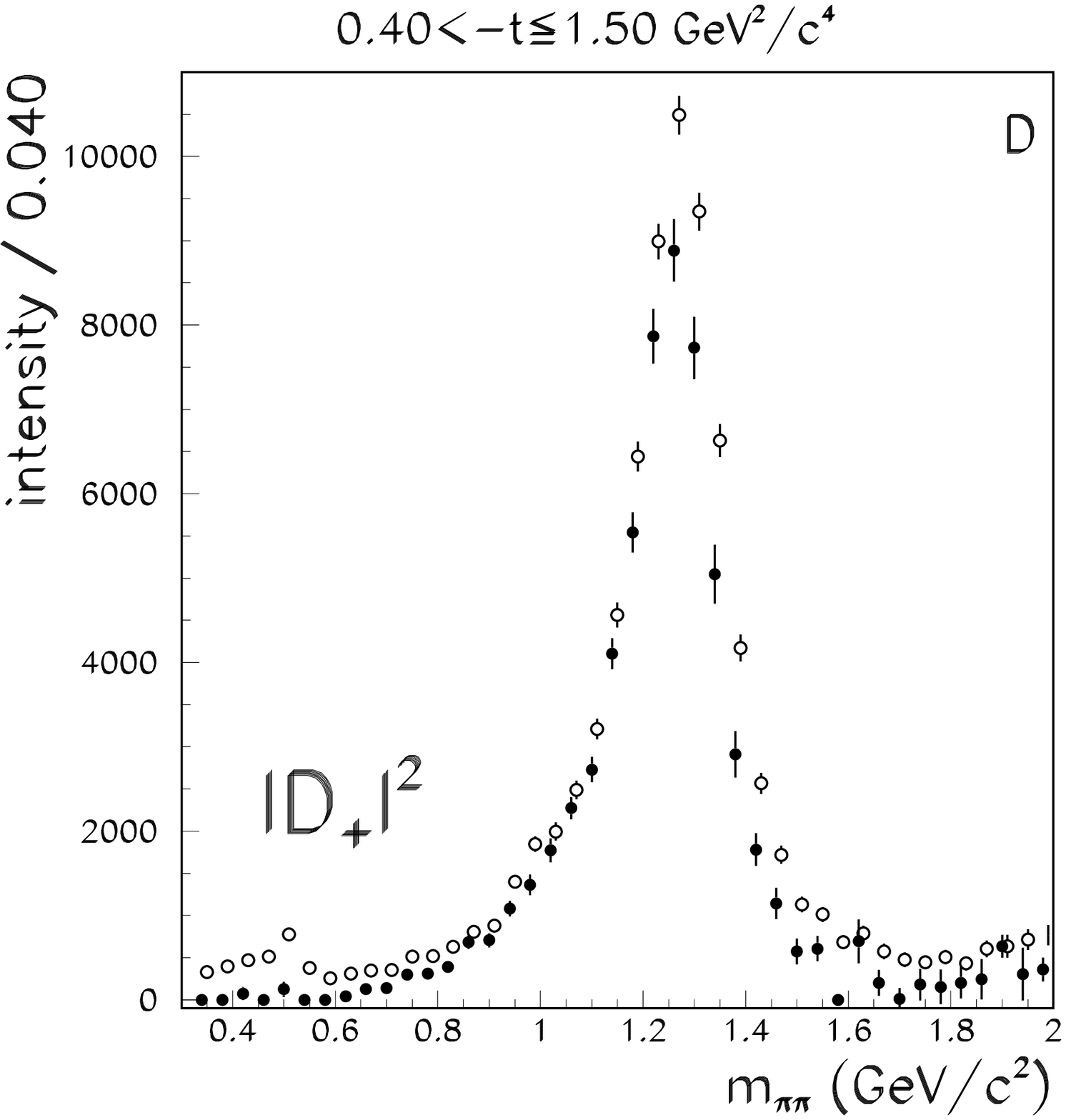,width=0.45\textwidth,height=0.45\textwidth}} \\ 
\mbox{\epsfig{file=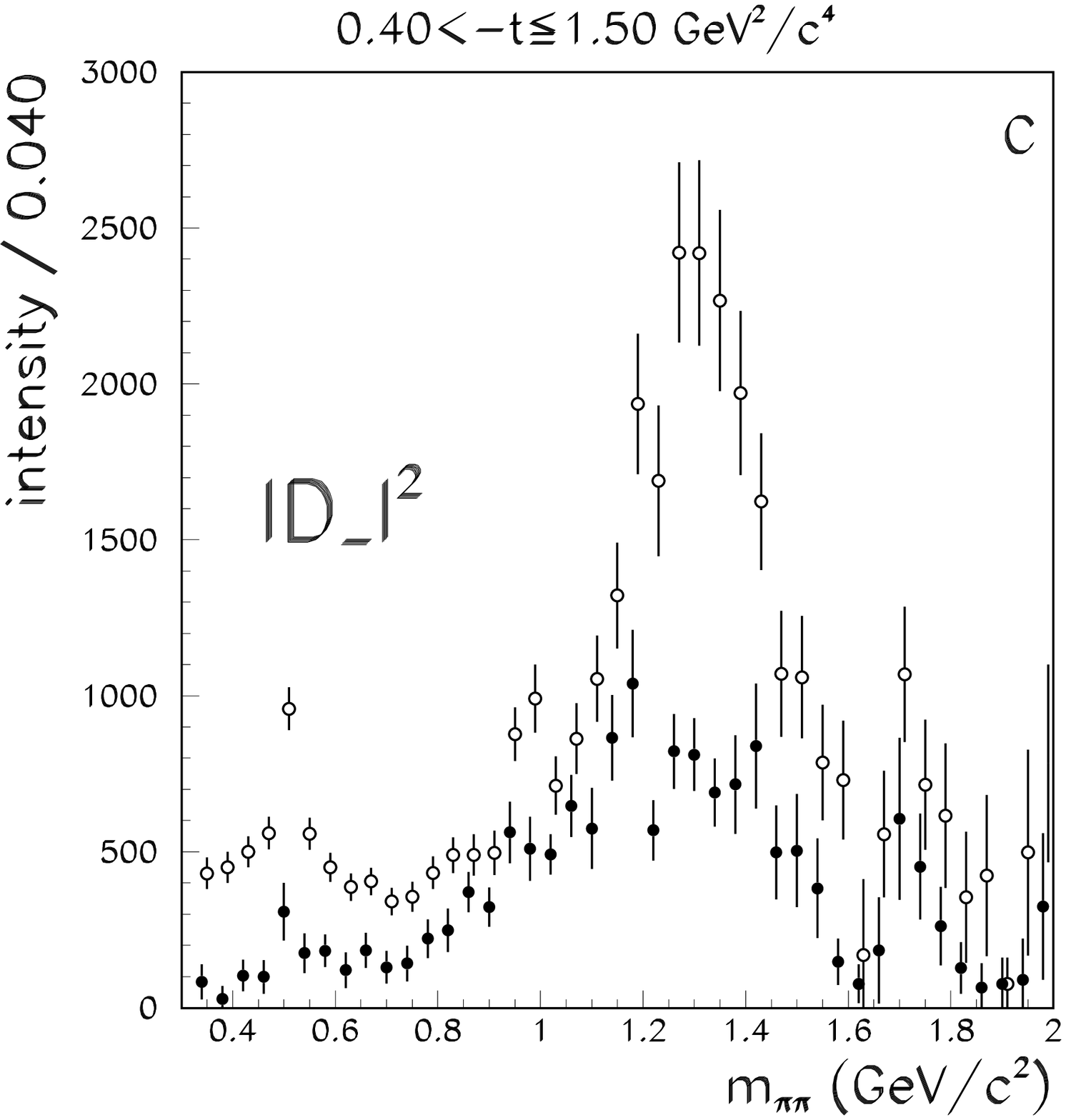,width=0.45\textwidth,height=0.45\textwidth}} &
\mbox{\epsfig{file=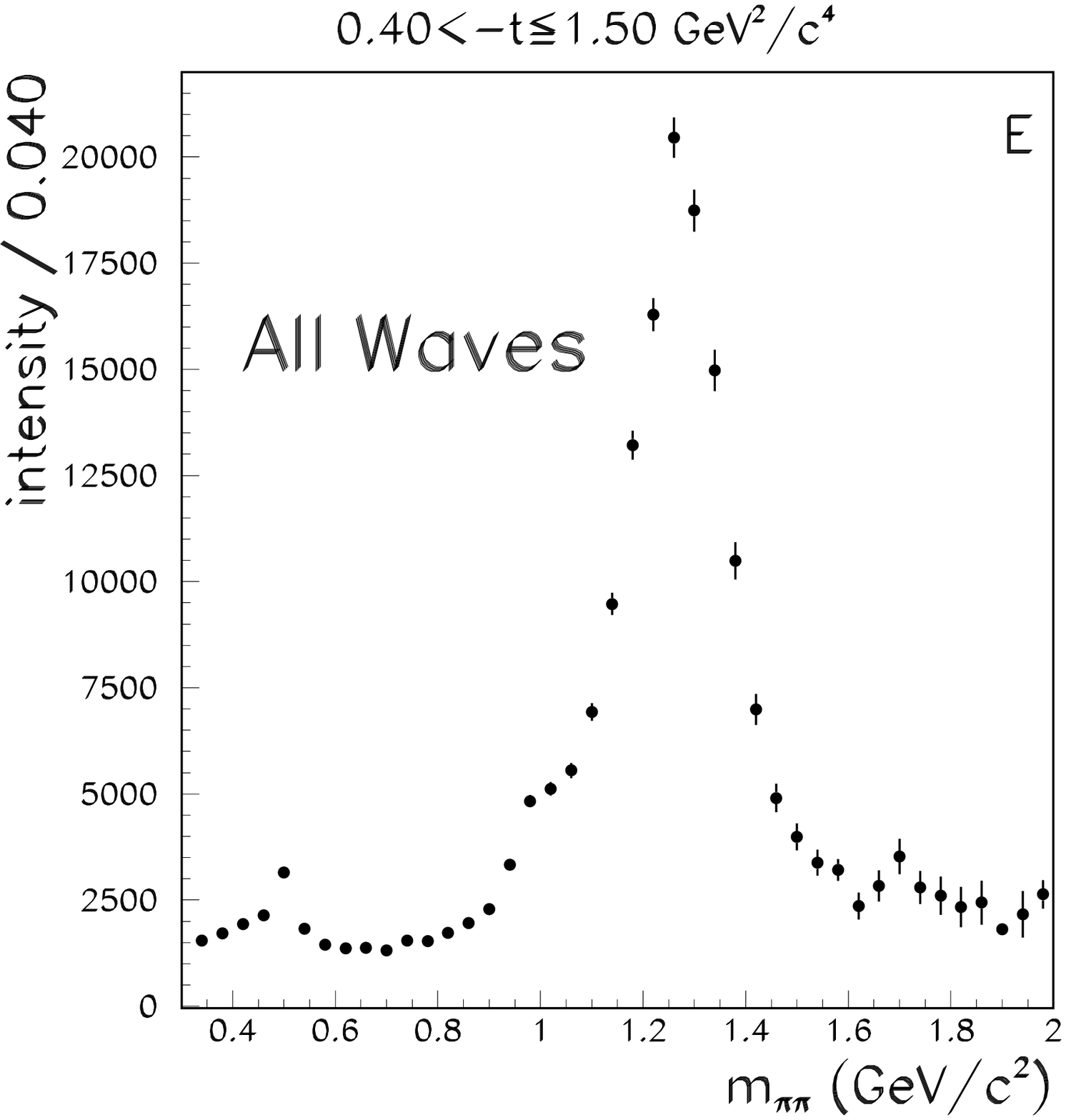,width=0.45\textwidth,height=0.45\textwidth}} 
\end{tabular}

\caption{\label{fig:t3intensities}The squares of the magnitudes of the partial 
waves (a)--(d) as a function of mass for events in the region 
$0.40<|t|<1.50\, GeV^{2}/c^{2}$.  The solid circles correspond to the
physical  solution. The
coherent sum of the partial waves integrated over decay angles, (e),  gives 
the acceptance corrected mass distribution.   The $D_+$ partial wave (natural parity
exchange) is the dominant partial wave.}  

\end{figure}

\begin{figure}[htbp]\centering
\begin{tabular}{cc}
\mbox{\epsfig{file=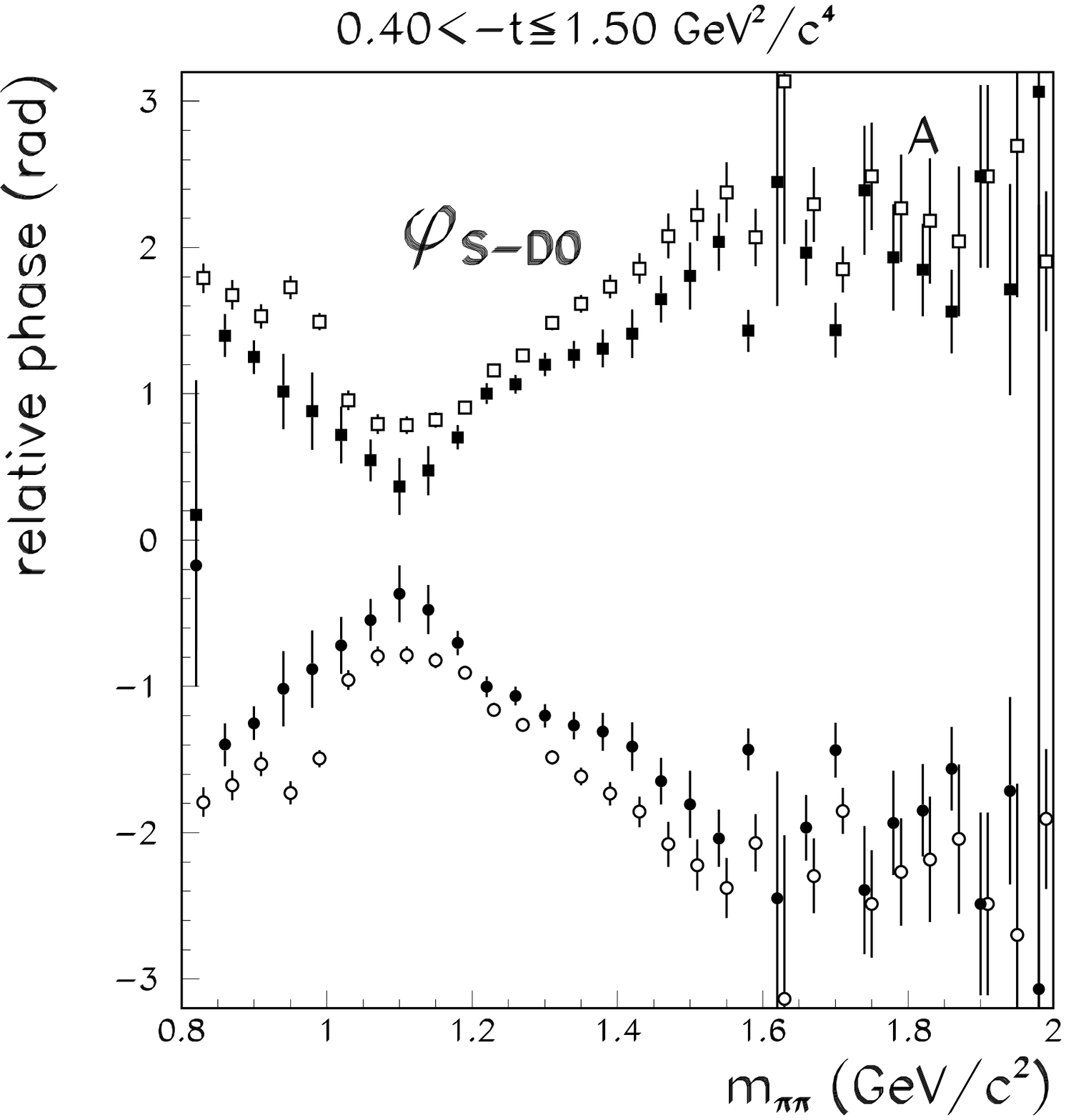,width=0.45\textwidth,height=0.45\textwidth}} &
\mbox{\epsfig{file=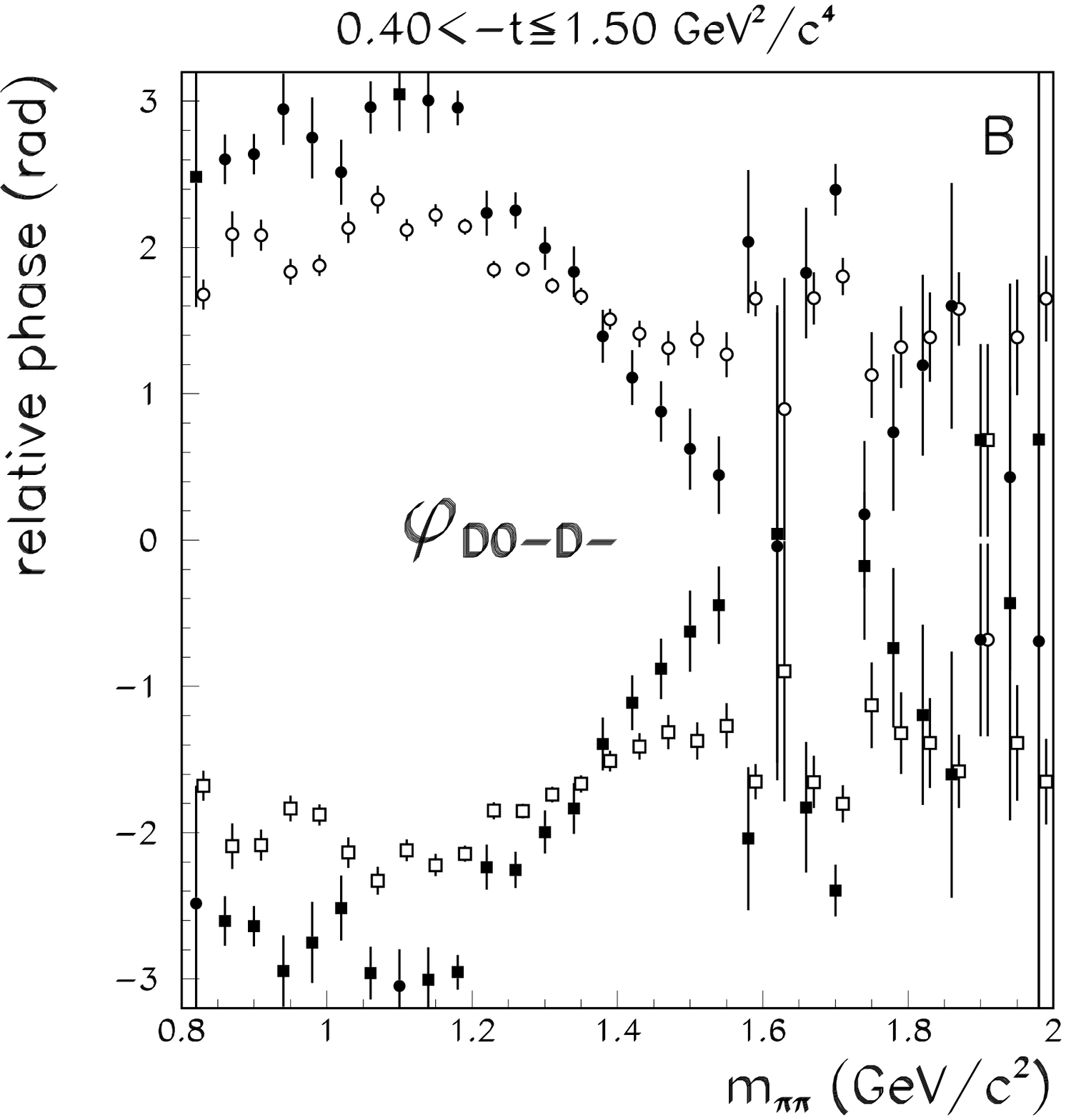,width=0.45\textwidth,height=0.45\textwidth}} \\
\end{tabular}

\caption{\label{fig:t3phases}The relative phases between unnatural parity exchange
partial waves for events in the region \protect$ 0.40<|t|<1.5$ .  
The physical solution (solid circles) in the $S-D_0$ relative phase plot 
(a) is more smoothly varying than in figures \ref{fig:t0phases} and \ref{fig:t1phases}. 
The $D_0-D_-$ relative phase (b) is constant only up to approximately $1.2 \ GeV/c^2$.}
\end{figure}

The partial wave analysis results in the region $ 0.40<-t<1.50\, GeV^{2}/c^{2} $
(Figures \ref{fig:t3intensities} and \ref{fig:t3phases}) are significantly
different from results at smaller $ |t| $. The bump observed in the mass plot 
(figure \ref{fig:mAsFofT}d) near
$ 1.0\, GeV/c^{2} $ is found in the $ S $-wave intensity.  The $ D_{+} $
partial wave is dominant (as opposed to the $D_0$ partial wave), indicating a shift from 
unnatural parity exchange
processes at small $ |t| $ to production via natural parity exchange at large
$ |t| $.

\subsubsection{Fine \protect$ |t|\protect $ Bin Fits}

The statistics of this experiment are sufficient to allow the region $ 0.00<-t<0.40\, 
GeV^{2}/c^{2} $
to be analyzed in finer $ |t| $ bins, nine in all,
 for masses up to approximately $1.8 \, GeV/c^2$. 
The $ |t| $ -dependence of the $ S $-wave intensity may be summarized by
noting that the ratio of the maxima in the intensities at approximately $ 0.8\, GeV/c^{2} $ 
and $ 1.3\, GeV/c^{2} $
decreases with increasing $ |t| $, and the ratio of height of maximum intensity at 
approximately $ 0.8\, GeV/c^{2} $ to value of
the intensity measured at $ 0.98\, GeV/c^{2} $ decreases.

The lineshape of the $ f_{2}(1270) $ in the $ D_{0} $-wave intensity
is largely independent of $ |t| $. The $ S-D_{0} $ relative phase is $ |t| $
dependent. 
The lineshape of the $ D_{+} $-wave is also independent of $ |t| $. More details
of the $ |t| $ dependence of the partial waves follow.

The intensities of the individual partial-waves and phase differences
as a function of mass for the nine bins in $ |t| $ for 
 $ 0.00<-t<0.40 \ GeV^{2}/c^{2} $ as well as for the $|t|$-bins 
presented in this paper are available on the World Wide Web \cite{WWW}.

\subsection{Model Dependent Fits of the \protect$ |t|\protect $- Distributions}

\begin{figure}[p]\centering
\begin{tabular}{cc}
\mbox{\epsfig{file=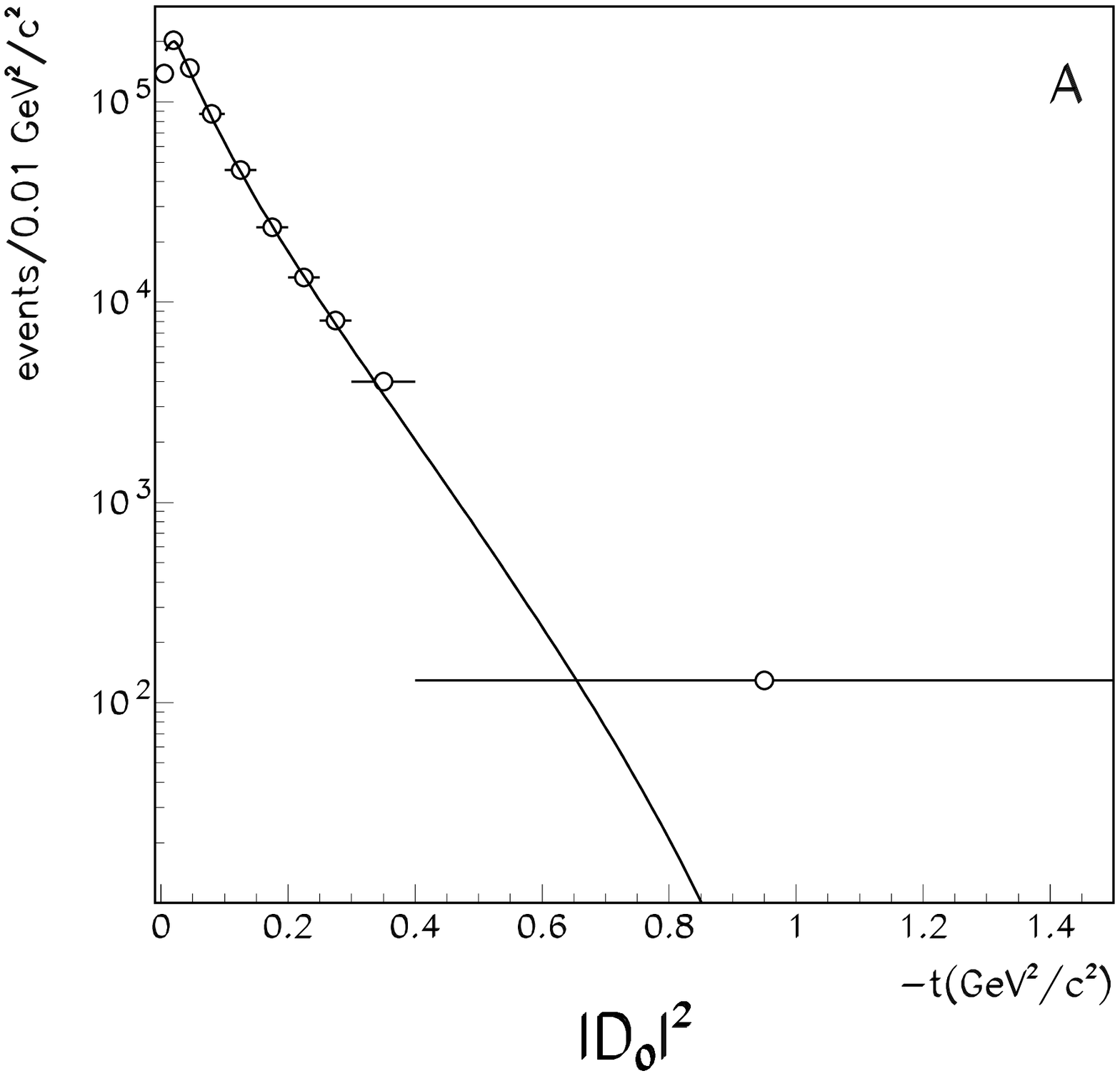,width=0.45\textwidth,height=0.45\textwidth}} &
\mbox{\epsfig{file=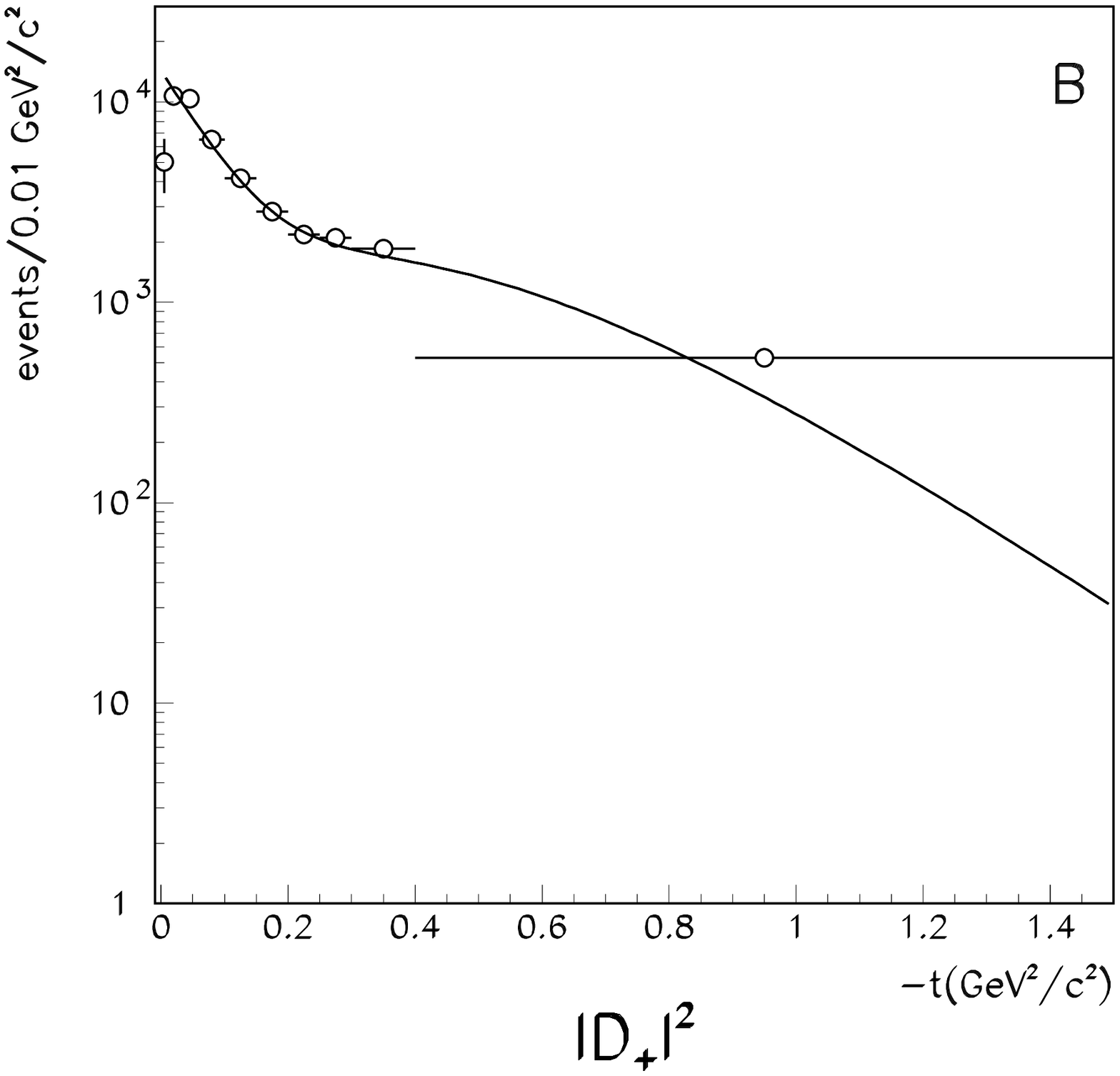,width=0.45\textwidth,height=0.45\textwidth}} \\
\end{tabular}

\caption{\label{fig:tdependentsummary2}The integrals of fitted 
relativistic Breit-Wigner functions 
over the peak regions of the $D_{0}$ (a) and $D_{+}$ (b) -waves as a function of
$|t|$ are fitted by one-pion-exchange and 
$a_2$-exchange with absorption 
as described in the text.}

\end{figure}

The integrals of fitted relativistic Breit-Wigner functions 
over the peak regions of the $D_{0}$ and $D_{+}$-waves as a function of $|t|$ are 
shown in Figure \ref{fig:tdependentsummary2}. 
The dependences of these intensities on $|t|$ are fitted to functions
given by Regge-exchange models. 
At low-$|t|$,  the unnatural 
parity
exchange $D_0$ partial wave is expected to be dominated by OPE. The Reggeized form 
for this contribution is given by

\begin{equation}
\frac{d|D_0|}{d|t|} = N_{D_0} |\sqrt{-t} e^{b_\pi t} (t-m^2_{f_2})^2
\left(1 + e^{i\pi\alpha(t)}\right)\Gamma(-\alpha_\pi(t))|^2,\;\;
\alpha_\pi(t) = 0.9(t-m_\pi^2) \label{eq:OPE}
\end{equation}

 In this expression, the $\sqrt{-t}$ factor is due to  helicity-flip
in the pion-nucleon coupling, and the polynomial dependence on $t$ arises from
the $f_2$ coupling to $\pi\pi$ at the production vertex. 
 The particular form of this dependence is due to the 
 angular momentum barrier factor proportional to $k^L$ with $L=2$ and 
 $k$ being the magnitude of the 3-momentum of the exchanged particle in
the $f_2$ rest
 frame (Gottfried-Jackson frame), given by 
$k^2=((m_{f_2}-m_\pi)^2-t)((m_{f_2}+m_\pi)^2 -t)/4m_{f_2}^2 \sim 
 (m^2_{f_2} - t)^2/4m^2_{f_2}$. 

 The slope, $b_\pi$, in the OPE form is $4.08\pm0.02/(GeV^2/c^2)$.
  The systematic uncertainty in the slope of the 
$\alpha_\pi(t)$ Regge trajectory is $\pm 0.1/(GeV^2/c^2)$. 
As shown by
 Irving and Michael \cite{IM} the natural parity exchange $D_+$ -wave is
 dominated by absorption of the pion exchange and may be parameterized in terms of a 
Regge cut in the nucleon helicity-flip amplitude

\begin{equation}
C = g_c e^{b_c t} e^{-{1\over 2}i \pi \alpha_C(t)}\left
({p_L\over p_0}\right)^{\alpha_C(t)-1}, \;\; \alpha_C(t) = 0.41 t,\;\; g_c = -0\
.84, \; b_c = 3.89
\end{equation}

 The  nucleon-flip and non-flip $a_2$
 exchange  is then given by,
\begin{equation}
A_f = g_a (-t) e^{b_a t} e^{-{1\over 2}i \pi \alpha_{A_2}(t)}
\left({p_L\over p_0}\right)^{\alpha_{A_2}(t)-1},\; \;
A_n = A_f {r \over {\sqrt{-t}}}
 \label{eq:a2excbegin}
\end{equation}
 respectively, with the parameters $\alpha_{A_2}(t) = 0.5 + 0.82 t$ and
 $g_a = 1.35$, $b_a = 3.24$, $p_0 = 17.2 \ GeV/c$, and $r=0.5$ from \cite{IM} and 
$p_L=18.3 \ GeV/c$, the beam momentum for these data.
 The $D_+$-wave intensity is then fitted to

\begin{equation}
\frac{d|D_+|^2}{d|t|} = N_{D_+} (|A_n|^2 + |A_f + C|^2) \label{eq:a2excend}
\end{equation}

 For both forms the fitted functions are averaged over the $ |t| $ bins shown in the plots.   
The plotted curves are calculated from the models without averaging.

\begin{figure}[htbp]\centering
\begin{tabular}{cc}
\mbox{\epsfig{file=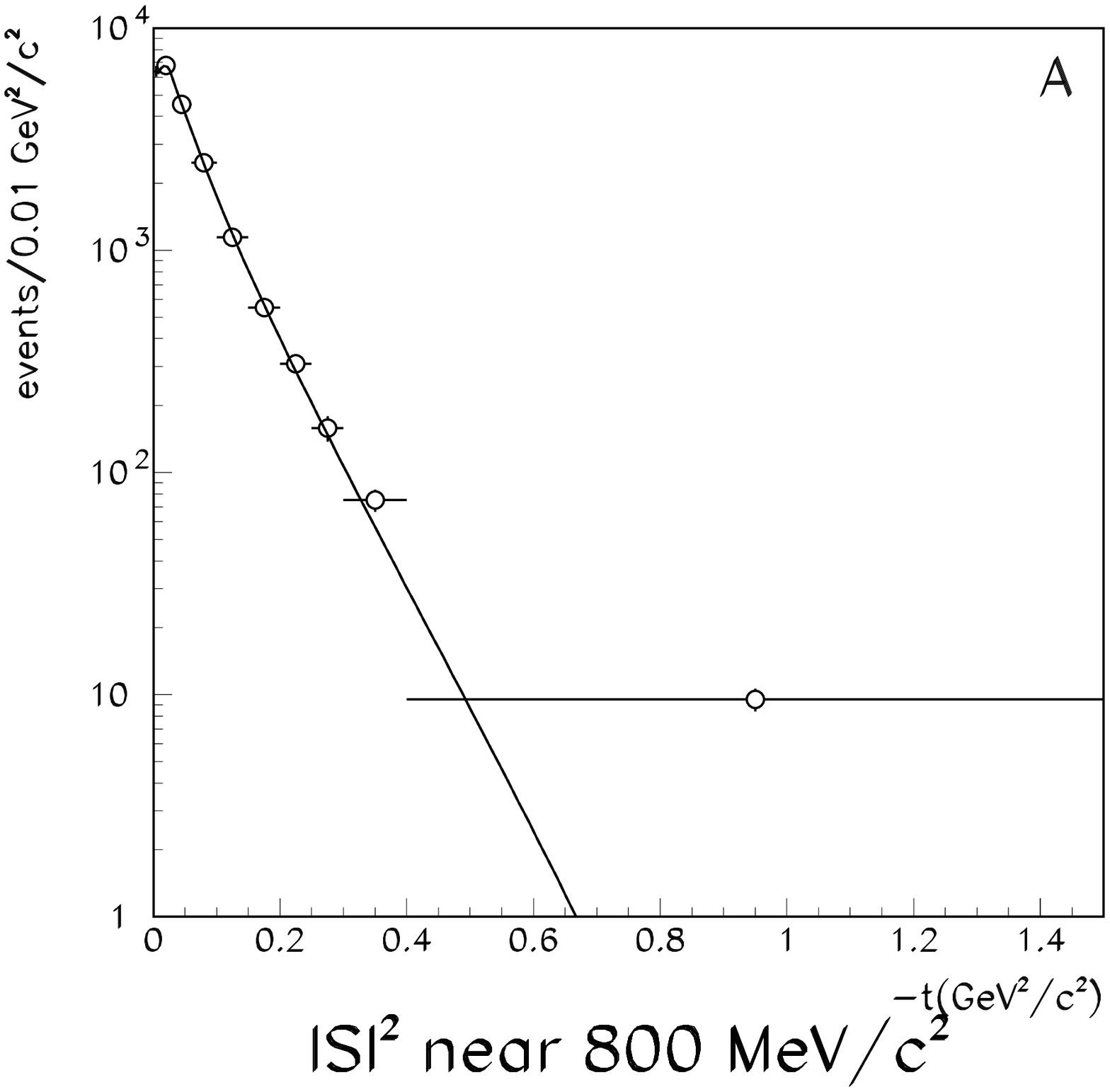,width=0.45\textwidth,height=0.45\textwidth}} &
\mbox{\epsfig{file=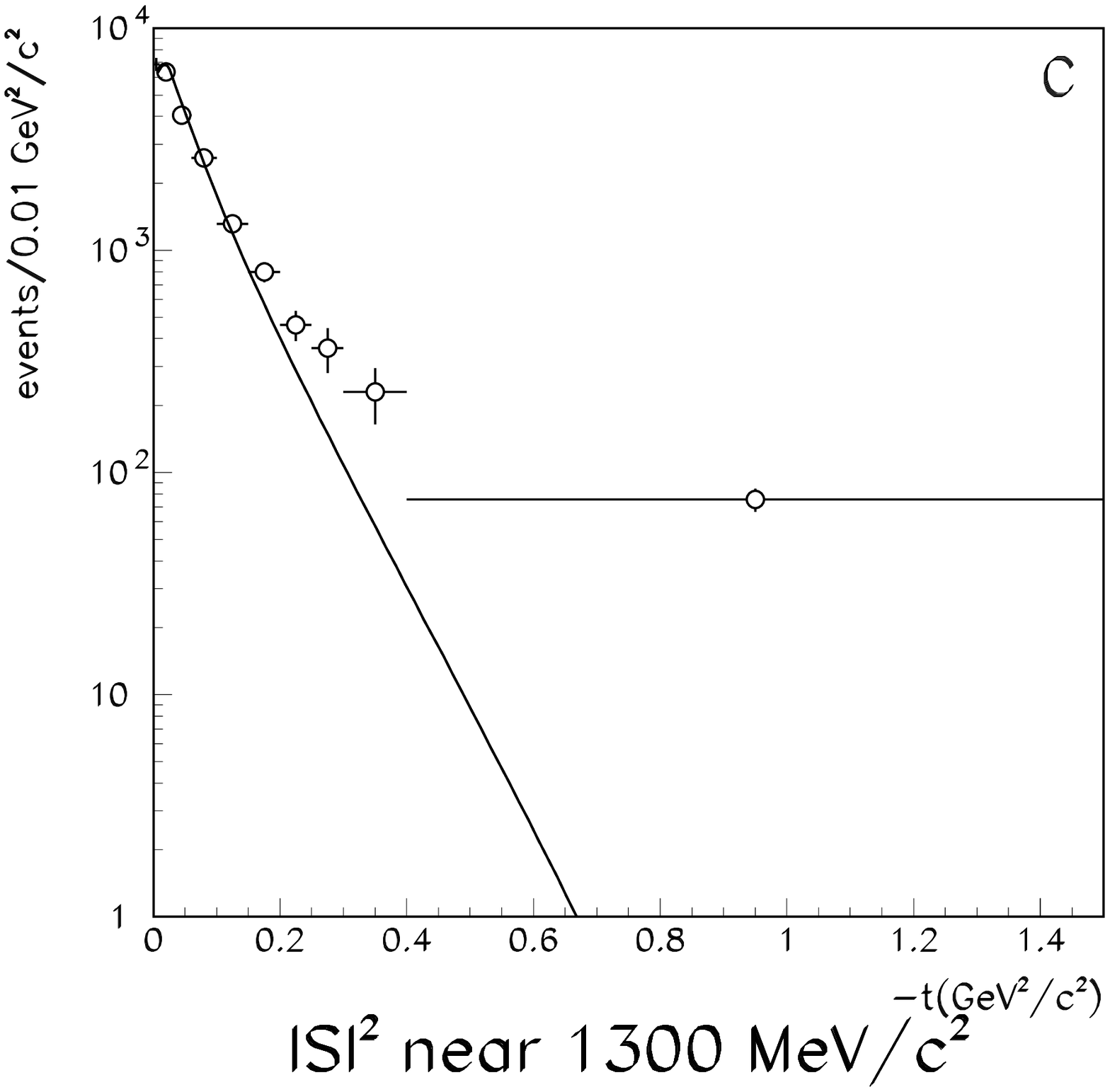,width=0.45\textwidth,height=0.45\textwidth}} \\
\mbox{\epsfig{file=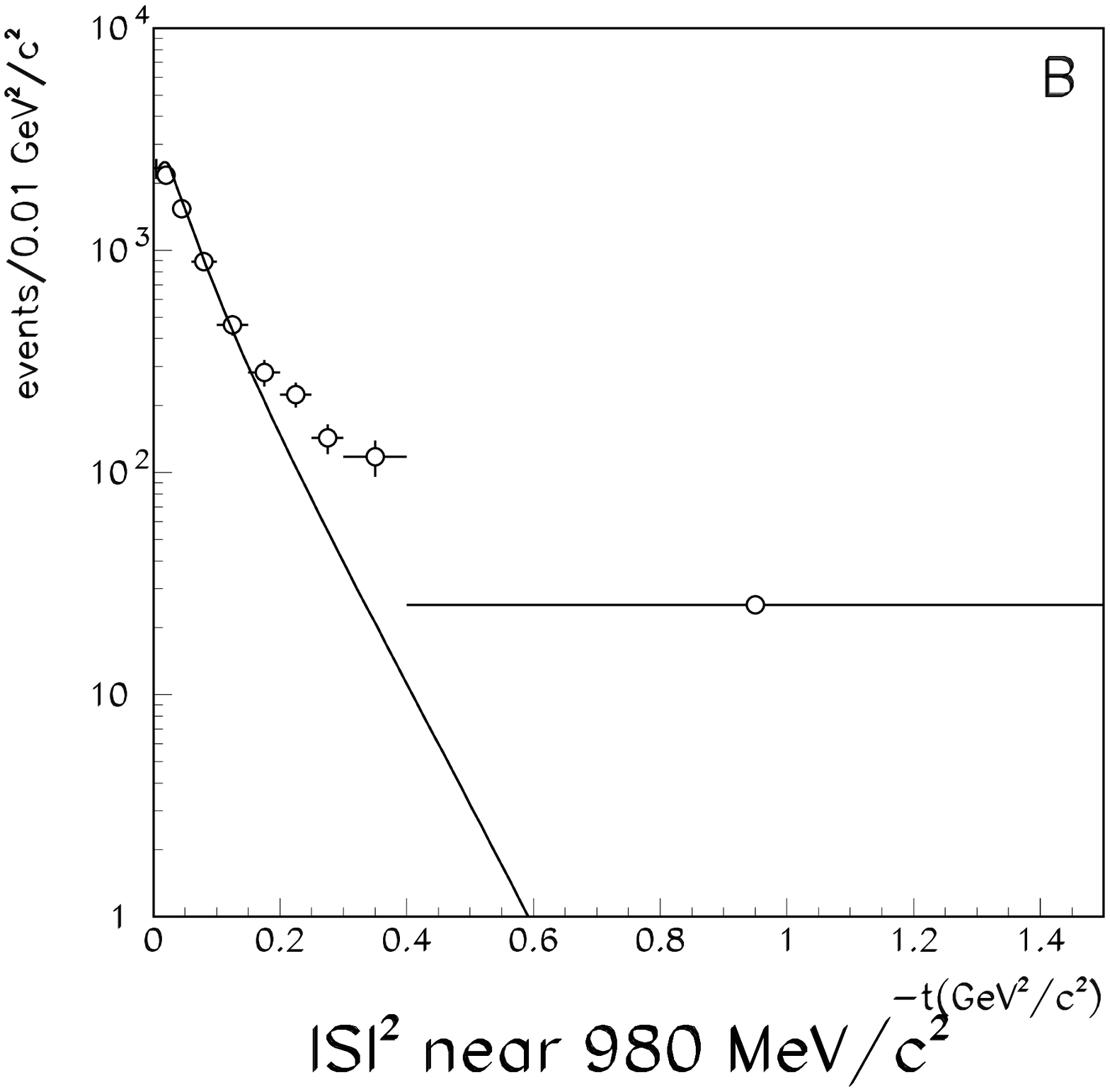,width=0.45\textwidth,height=0.45\textwidth}} \\
\end{tabular}

\caption{\label{fig:tdependentsummary}The \protect$ S\protect $-wave intensities
at three different masses ($0.80,0.98,$ and $1.30 \ GeV/c^2$) are compared with a one-pion 
exchange form.  Except for the overall normalization, the parameters of the 
OPE parameterization are those determined in the fit to figure \ref{fig:tdependentsummary2}(a).  
OPE describes the data well at small $|t|$.  
The excess of events at higher $|t|$ in (b) and (c) is consistent with the
existence of additional production mechanisms that are less strongly biased toward 
small momentum-transfer-squared production than is OPE. }

\end{figure}

In figure \ref{fig:tdependentsummary} the peak value of the $ S $-wave intensity near $ 
0.80\, GeV/c^{2} $,
the value of the $ S $-wave intensity at $ 0.98\, GeV/c^{2} $ and the peak value
of the $ S $-wave intensity at approximately $ 1.3\, GeV/c^{2} $ as a function of $ |t| $ 
are shown. A  one-pion-exchange form similar to Equation \ref{eq:OPE}, but with the 
  $t-m^2_{f_2}$ factor removed, is overlayed on these distributions.   The Regge 
trajectory slope and exponential slope are fixed to the values found for the  $D_0$-wave fit, 
and a one parameter fit is used to set the normalization. At small values of  $|t|$ the OPE 
form qualitatively agrees with the data. 
The excess of events at higher $|t|$ in (b) and (c) is consistent with the
existence of additional production mechanisms that are less strongly biased toward 
small momentum-transfer-squared production than is OPE.

\section{Conclusions}

A partial wave analysis was carried out on a sample of 847,460 events of
the reaction $ \pi ^{-}p\rightarrow \pi ^{0}\pi ^{0}n $ collected by experiment
E852. The PWA was performed in $ 0.04\, GeV/c^{2} $-wide
bins in di-pion mass ($m_{\pi ^0\pi ^0}$)
and momentum-transfer-squared ($|t|$) from the incident $\pi^-$ to the outgoing
$\pi^{0}\pi^{0}$ system.  Coarse and fine binning in $|t|$ were used.
Numerical values for the partial wave intensities and phases
as a function of di-pion mass, for coarse and fine bins in $|t|$,
are available on the World Wide Web \cite{WWW}.
 The $ f_{2}(1270) $ meson is found to be produced by unnatural parity
exchange at small values of $|t|$ and natural parity exchange at
large values of $|t|$. The $|t|$ dependences of $D_0$-wave and 
$D_+$-wave intensities 
are consistent with 
Regge-exchange models.  An enhancement in the
$G_{0}$ wave consistent with the $ f_{4}(2050) $ meson is observed in unnatural
parity exchange at small momentum transfer. The shape of the $ S $-wave intensity
has a strong momentum transfer dependence. The presence of dips in the $ S $-wave
intensity near $ 0.98 $ and $ 1.5\,GeV/c^{2} $, accompanied by rapid phase
variations relative to the $ D_{0} $-wave is consistent with similar 
observations reported in reference \cite{Ald98} and in centrally produced
$\pi^{0}\pi^{0}$ systems in 450 $GeV/c$ $pp$ collisions \cite{Barb}.  The
latter claims evidence for the $f_{0}(980)$ and $f_{0}(1500)$.
At large momentum transfer,
the $ f_{0}(980) $ meson is observed as a bump in the $ S $-wave intensity.
The $S$-wave intensity in the peak near $0.80 \ GeV/c^2$ is well-described by OPE. 
It should also be noted that the model of Anisovich {\it et al} 
\cite{Ani96,Ani97,Ani97b} predicts the presence of a dip in the $|t|$ distribution for this 
mass region near $|t|\approx 0.07 GeV^2/c^2$ \cite{Achasov}.  In direct contradiction, 
no such dip is observed in this analysis.  At higher  masses the $S$-wave is adequately 
described by OPE only at small values of $|t|$.

\section{Acknowledgements}
We wish to thank the members of the MPS group at BNL as well as the staffs
of the AGS, of BNL and of the various colaborating institutions.  This
work was supported in part by the U. S. Department of Energy, the National
Science Foundation and the Russian State Committee for Science and
Technology.


\pagebreak

\pagebreak

\pagebreak

\pagebreak

\begin{table}
{\centering \begin{tabular}{|c|c|c|c|}
\hline 
Partial wave&
L&
$|m|$&
Naturality of the exchange particle\\
\hline 
\hline 
$ S $&
0&
0&
unnatural\\
\hline 
$ D_{0} $&
2&
0&
unnatural\\
\hline 
$ D_{-} $&
2&
1&
unnatural\\
\hline 
$ G_{0} $&
4&
0&
unnatural\\
\hline 
$ D_{+} $&
2&
1&
natural\\
\hline 
\end{tabular}\par}

\caption{\label{tab:nomenclature}The nomenclature for partial waves includes 
the angular momentum (\protect$ L\protect $) of the 	
\protect$ \pi ^{0}\pi ^{0}\protect $
system, the magnitude of the magnetic quantum number 
(\protect$ m\protect $)
and the naturality of the exchange particle which leads to production in the
particular partial wave. The naturality is natural if 
\protect$ P = (-1)^{J}\protect$ and unnatural if
\protect$ P = (-1)^{J+1}\protect$.}
\end{table}

\end{document}